\documentclass[journal]{IEEEtran}

\IEEEoverridecommandlockouts
\makeatletter
\def\ps@headings{%
	\def\@oddhead{\mbox{}\scriptsize\rightmark \hfil \thepage}%
	\def\@evenhead{\scriptsize\thepage \hfil \leftmark\mbox{}}%
	\def\@oddfoot{}%
	\def\@evenfoot{}}
\makeatother
\usepackage{subfigure}
\usepackage{colortbl}
\usepackage{bm}

\usepackage{graphicx}  
\usepackage{url}       

\usepackage{amsmath}   
\usepackage{cite}

\usepackage{amsfonts,amssymb}
\usepackage{cite}
\usepackage{stfloats}

\usepackage{cases}
\usepackage{algorithm}
\usepackage{multirow}
\usepackage{algorithmic}
\usepackage{stfloats}
\usepackage{epstopdf}

\makeatletter

\newcommand{\Rmnum}[1]{\expandafter\@slowromancap\romannumeral #1@}
\makeatother

\UseRawInputEncoding

\newtheorem{proposition}{{Proposition}}

\newtheorem{corollary}{{Corollary}}

\newcommand{\ls}[1]
{\dimen0=\fontdimen6\the\font
	\lineskip=#1\dimen0
	\advance\lineskip.5\fontdimen5\the\font
	\advance\lineskip-\dimen0
	\lineskiplimit=.9\lineskip
	\baselineskip=\lineskip
	\advance\baselineskip\dimen0
	\normallineskip\lineskip
	\normallineskiplimit\lineskiplimit
	\normalbaselineskip\baselineskip
	\ignorespaces
}

\begin{document}
	\title{Towards Reliable Emergency Wireless Communications over SAGINs: A Composite Fading and QoS-Centric Perspective}
	\vspace{10pt}
\author{	\IEEEauthorblockN{Yinong Chen, \emph{Student Member}, \emph{IEEE}, Wenchi Cheng, \emph{Senior Member}, \emph{IEEE}, Jingqing Wang, \emph{Member}, \emph{IEEE},\\Xiao Zheng, \emph{Student Member}, \emph{IEEE}, and Jiangzhou Wang, \emph{Fellow}, \emph{IEEE}}\vspace{-17pt}
		\thanks{
		Yinong Chen, Wenchi Cheng, Jingqing Wang, and Xiao Zheng are with the State Key Laboratory of Integrated Services Networks, Xidian University, Xi'an, 710071, China (e-mails: yinongchen@stu.xidian.edu.cn;  wccheng@xidian.edu.cn; jqwangxd@xidian.edu.cn; zheng\_xiao@stu.xidian.edu.cn). 
		
		Jiangzhou Wang is with the School of Information Science and Engineering, Southeast University, and Purple Mountain Laboratories, Nanjing 211119, China (e-mail: j.z.wang@kent.ac.uk).}
	}
		
		

	%

	\maketitle

	\begin{abstract}
	In emergency wireless communications (EWC) scenarios, ensuring reliable, flexible, and high-rate transmission while simultaneously maintaining seamless coverage and rapid response capabilities presents a critical technical challenge. 
	To this end, satellite-aerial-ground integrated network (SAGIN) has emerged as a promising solution due to its comprehensive three-dimensional coverage and capability to meet stringent, multi-faceted quality-of-service (QoS) requirements. 
	Nevertheless, most existing studies either neglected the inherent characteristics of the complex channel conditions due to the terrain changes or analyzed the performance in the absence of QoS constraints, resulting in a mismatch between theoretical analysis and practical performance. To remedy such deficiencies, in this paper we establish a performance modeling framework for SAGIN  employing the Fisher-Snedecor $\mathcal{F}$ composite fading model to characterize the air-ground link.
	In specific, the proposed $\mathcal{F}$ composite fading channel is adopted to accurately describe both multipath fading and shadowing in harsh ground environments. The exact distribution of end-to-end signal-to-noise (SNR) statistics for space-air and air-ground links is developed, enabling theoretical analysis of cascaded channels with fixed-gain amplify-and-forward (AF) and decode-and-forward (DF) relaying protocols, respectively. 
	Furthermore, asymptotic expressions of the derived results are provided to offer concise representations and demonstrate close alignment with theoretical predictions in the high-SNR regime. Finally, the insightful closed-form and asymptotic expressions of effective capacity with QoS provisioning, outage probability, and $\epsilon$-outage capacity are investigated, respectively, followed by both field measurements and Monte Carlo simulations to verify the effectiveness.
    \end{abstract}

\section{Introduction}
\IEEEPARstart{R}{eliable} and flexible wireless communications with rapid response capability is a key point to guarantee public safety during various natural disasters\cite{EWCUAV}. 
In the event of natural disasters such as earthquakes, floods and typhoons, critical communication infrastructure is often damaged or completely destroyed under such challenging environments, severely hindering timely information exchange and coordination. 
One of the promising solutions is to quickly deploy terrestrial emergency vehicles, which act as temporary base stations (BSs) in the affected areas to support data dissemination with acceptable overhead. However, these traditional ground-based approaches often  face challenges of reaching the disaster area promptly due to terrain obstructions, damaged roads, or unsafe conditions in the disaster zone, which suffer from limited accessibility and exhibit difficulty to achieve dynamic rescue demands\cite{Harsh}. 

In light of this issue, unmanned aerial vehicles (UAVs) have emerged as a promising solution by acting as flying BSs and/or as relays to rapidly reestablish communication links. Their flexibility, mobility, and ease of deployment make UAVs a powerful enabler of emergency wireless communication (EWC), particularly in time-critical scenarios such as search and rescue missions\cite{ScienceChina, Yao}.
To date, considerable research efforts have been devoted to the statistical performance characterization and optimization of UAV-assisted wireless networks\cite{UAVBS1,UAVrelay1,UAVrelay2}. 
Among various applications, UAVs equipped with communication payloads have demonstrated great potential in supporting cooperative relaying.
Depending on the signal processing complexity and application needs, two representative relaying schemes, comprising amplify-and-forward (AF) and decode-and-forward (DF), have been widely studied in the context of UAV-based communications.
For example, the authors of \cite{UAVrelay1} analyzed the outage probability of AF relay-based UAV system under both line-of-sight (LoS) and non line-of-sight (NLoS) connections modeled by Rician and Rayleigh fading channels, respectively. 
\cite{UAVrelay2} analyzed the secrecy performance of a DF relay-based UAV system in urban environments, evaluating secrecy performance under various factors, including UAV altitude, UAV transmit power, and urban environment types.
Nevertheless, UAVs also have some typical shortcomings, such as constrained energy and payload capacity, as well as unclear channel model in dynamic environments, imposing performance bottlenecks in terms of capacity, energy efficiency, and coverage\cite{UAVshorts}.

To overcome the inherent limitations of traditional UAV-based network, space-air-ground integrated network (SAGIN) has emerged as a promising paradigm. 
By integrating aerial and spaceborne nodes, SAGIN introduces an additional altitude dimension into network design, thereby extending coverage, improving link reliability, and enabling high-data-rate communications over a significantly broader area compared to purely aerial or terrestrial networks \cite{SAGIN_intro1,SAGIN_intro2,SAGIN_intro3,SAGIN-NAKAM,saginshannon}.
Generally, UAVs play a critical role as aerial relay nodes in SAGIN, bridging ground and space networks. 
Considering the multi-hop channel, to harness the superiority of SAGIN, some existing works explored its potentials via performance analysis, revealing the interaction among cooperative channels including the space-air, space-ground and air-ground links. 
The authors in \cite{SAGIN1} took the realistic propagation environment, high-altitude platform (HAP) mobility, and mathematical tractability into account and reconstructed the cooperative channel models for SAGIN, with which the outage probability and asymptotic outage probability in closed-form were given. The authors in \cite{SAGIN2} calculated the UAV energy outage probability and signal-to-noise ratio (SNR) outage probability for SAGIN uplink transmission with an integrated model that considered UAV channel fading, energy consumption, and harvesting. The authors in \cite{SAGIN3} evaluated the outage performance of SAGIN in the 3D Poisson field and approximated its outage probability in closed-form expression by utilizing homogeneous Poisson Point Process. \cite{SAGIN4} provided a detailed analysis of the different distributions of the serving and interfering platforms, followed by exact and closed-form expressions via stochastic geometry. 

To completely harness the superiority of SAGIN in EWC scenarios, new challenges must to be addressed. 
On one hand, due to the terrain changes caused by secondary disasters, communication channels will experience severe multipath fading and shadowing under such a complex and dynamic propagation environment, which necessitates accurate and adaptable channel model to ensure reliable system performance.
Despite this, a series of existing research focuses on the Rayleigh,  Rician, and Nakagami-${m}$ fading\cite{SAGIN1,SAGIN2,SAGIN3,saginshannon,SAGIN-NAKAM} to model the air-ground links in SAGIN.  However, these models fall short in capturing the composite fading behavior in realistic post-disaster environments, where both multipath and shadowing effects are significant. 

On the other hand, time-sensitive and resource-constrained EWC scenarios demand strict consideration of quality-of-service (QoS) constraints. While Shannon capacity has been widely used in existing works \cite{SAGIN_intro3,SAGIN-NAKAM,saginshannon}, it only offers an upper bound under ideal, delay-unconstrained conditions. To better reflect practical  limitations, effective capacity has emerged as a meaningful alternative, quantifying the maximum constant arrival rate under statistical QoS constraints \cite{EC1,EC2}. This makes it particularly suitable for EWC applications requiring low-latency and guaranteed reliability.
Moreover, as SAGIN evolves into a heterogeneous and hierarchical architecture, classical metrics fall short in capturing differentiated QoS demands. Existing models often lack the flexibility needed to support diverse service requirements, limiting the network’s scalability and adaptability in emergency conditions. Therefore, developing statistical, QoS-aware expressions for effective capacity under delay and resource constraints is essential for enabling practical, real-time performance evaluation in SAGIN-based EWC systems.


Towards this end, with lower complexity and better fit to real channel conditions, the Fisher-Snedecor $\mathcal{F}$ distribution can more precisely reveal the characteristics of wireless channels by embedding with multipath fading and shadowing effects\cite{myf,Fchannel,Fchannel3}. It is noteworthy that \cite{tcomF} introduced a novel modified Fisher-Snedecor $\mathcal{F}$ fading channel model that resolves the mismatch between the mean power in existing distribution equations and the statistical result of the mean power under harsh shadowing conditions. 
This paper investigates a performance modeling framework for emergency SAGIN systems. To address the limitations of conventional fading models under post-disaster scenarios, we adopt the modified Fisher-Snedecor $\mathcal{F}$ distribution to accurately describe both multipath fading and shadowing of air-ground link. Meanwhile, the space-air link is modeled with the shadowed-Rician distribution to incorporate additional large-scale fading effects.
Based on this composite channel setup, we derive exact and asymptotic expressions for the end-to-end SNR statistics, including space-air link, air-ground link, and the cascaded links with both fixed-gain AF and DF relaying protocols. These results enable closed-form analysis of key performance metrics, including effective capacity under QoS constraints, outage probability, and $\epsilon$-outage capacity. Moreover, we validate the superiority of the proposed $\mathcal{F}$-based channel model through field measurements, demonstrating its improved fitting accuracy over  Rayleigh and Nakagami-$m$ models in post-disaster scenarios, followed by Monte Carlo simulations to verify the consistency between analytical results and simulations.

The rest of this paper is organized as follows. Section \ref{sec2} describes the SAGIN system model under the EWC scenario. Section \ref{sec3} characterizes the exact and asymptotic distribution of the end-to-end SNR statistics. Section \ref{sec4} presents the expressions for effective capacity, outage probability, and $\epsilon$-outage capacity. Section \ref{sec5} presents field measurements of the proposed Fisher-Snedecor $\mathcal{F}$ channel and Monte Carlo simulations to verify the accuracy and practical relevance of our analytical framework. The paper concludes with Section \ref{sec6}.

\section{System Model}\label{sec2}

\begin{figure}
	\centering
	\includegraphics[width=.96\columnwidth]{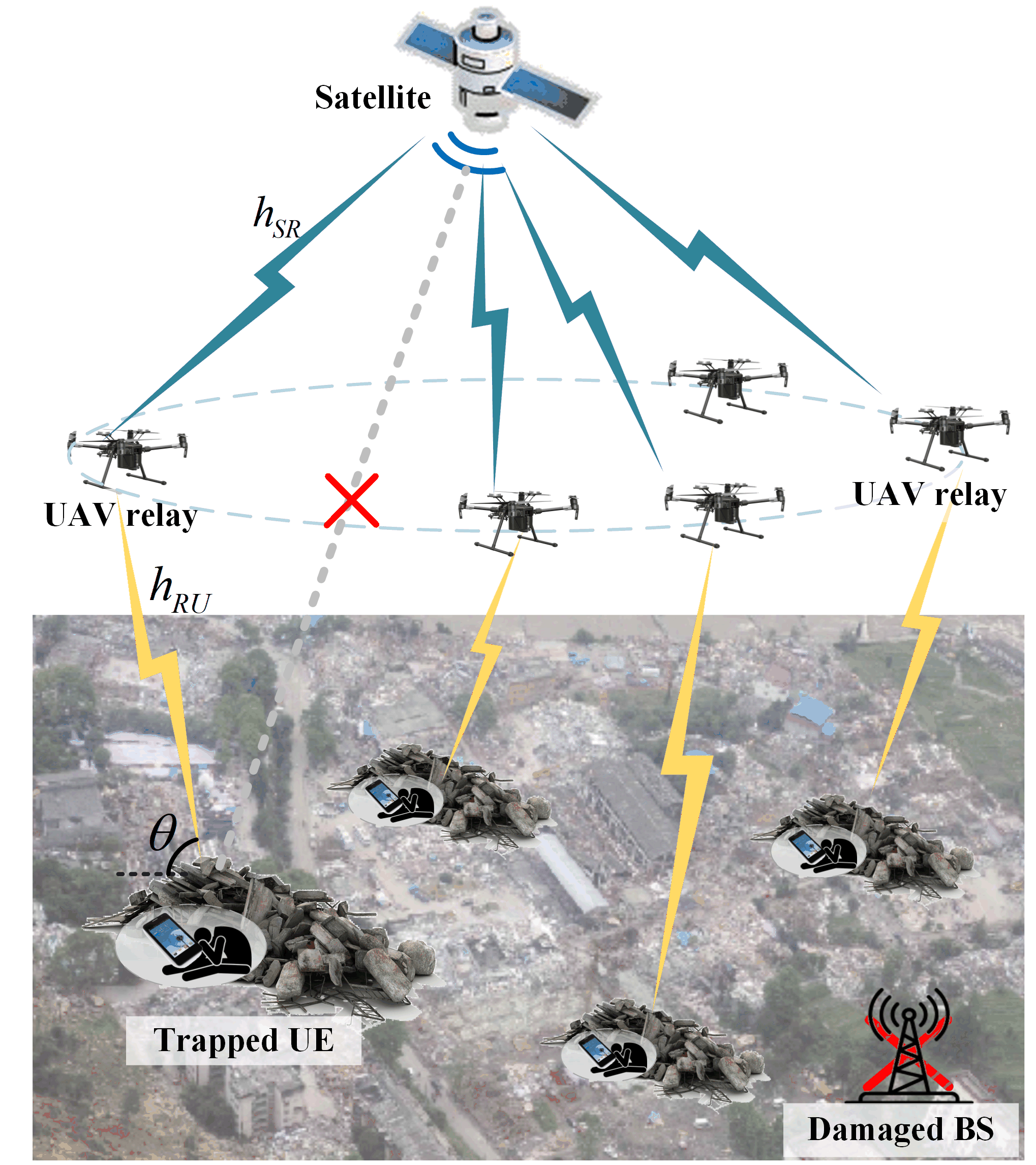}\vspace{-2pt}
	\caption{The considerd EWC system model with SAGIN.}
	\label{Fig_model}\vspace{-2pt}
\end{figure}

The proposed architecture for post-disaster scenarios is depicted in Fig.~\ref{Fig_model}, where the ground BSs are damaged and users are trapped beneath collapsed ruins. Given the obstruction of direct satellite-to-user links equipment (UE) caused by the complex terrain changes and debris with secondary disasters, the connection is reconstructed through UAV relays, establishing dual-hop connectivity via: 1) A satellite-to-UAV relay (S-R) link and 2) A UAV relay-to-UE (R-U) link.
For simplicity, we assume that all the communication nodes are deployed with one antenna. Without loss of generality, it is also assumed that the satellite and UAV relay operate within the same frequency band (L band or S band). To provide a more comprehensive analysis, both fixed-gain AF and DF relaying protocols for UAV relays are considered in this paper. 
Crucially, the heterogeneous nature of SAGIN induces different channel characteristics in each link segment, influenced by different physical mechanisms. This fundamental disparity necessitates layered channel analysis for accurate end-to-end performance modeling.
\subsection{Satellite-UAV relay (S-R) link}

The shadowed-Rician fading model presented in \cite{shadowed1,shadowed2,shadowed3,shadowed4} effectively captures the unique characteristics of satellite channels, including both line-of-sight (LoS) components and the effects of shadowing caused by obstructions such as buildings, foliage, and atmospheric phenomena. This model has been shown to provide a high degree of accuracy and practicality across various frequency bands. Therefore, the channel between the satellite and UAV relay, denoted by $h_\text{SR}$, adopts shadowed-Rician fading model. The probability density function (PDF) of the corresponding channel power gain, denoted by $f_{\left|h_{_\text{SR}}\right|^{2}}(x)$, can be expressed as follows:
\begin{equation}
	f_{\left|h_{_\text{SR}}\right|^{2}}(x)=\alpha_{_\text{SR}} e^{-\beta_{_\text{SR}} x}{ }_{1} F_{1}\left(\Gamma_{s}, 1, \delta_{_\text{SR}} x\right), x>0,\label{pdf-sr-channel}
\end{equation}	
where 
\begin{align}
	\begin{array}{l}
		\alpha_{_\text{SR}}=\frac{1}{2 b_{s}}\left[\frac{2 b_{s} \Gamma_{s}}{2 b_{s} \Gamma_{s}+\Omega_{s}}\right]^{\Gamma_{s}} ; \\
		\beta_{_\text{SR}}=\frac{1}{2 b_{s}} ; \\
		\delta_{_\text{SR}}=\frac{\Omega_{sr}}{2 b_{s}\left[2 b_{s} \Gamma_{s}+\Omega_{sr}\right]}.
	\end{array}
\end{align}
Therein, $\Omega_{sr}$ denote the average power of the LoS component, $2 b_{s}$ is the average power of the multi-path component, $\Gamma_{s} \in [0,\infty)$ denotes the Nakagami-$m$ parameter, and ${ }_{1} F_{1}\left(\Gamma_{s}, 1, \delta_{_\text{SR}} x\right)$ represents the confluent hypergeometric function of the first kind\cite[Eq. (9.14.1)]{2007Table}, which can be expressed as follows:
\begin{equation}
	{ }_{1} F_{1}\left(\Gamma_{s}, 1, \delta_{_\text{SR}} x\right)=e^{\delta_{_\text{SR}} x}\sum_{k=0}^{\Gamma_{s}-1}\frac{(-1)^k(1-\Gamma_{s})_{(k)}(\delta_{_\text{SR}} x)^k}{(k{!})^2},
\end{equation}
where $(\cdot)_{(k)}$ denotes the Pochhammer symbol. 

Since the free space transmission dominates the S-R channel, the path loss between the satellite and the UAV-relay, denoted by $PL_{_\text{SR}}$, can be expressed by
\begin{equation}
	PL_{_\text{SR}}=  \frac{\lambda}{4 \pi \left(d_{_\text{SR}}\right)^{\alpha_p}}, 
\end{equation}
where $\alpha_p$ denotes the path loss exponent generally satisfying $\alpha_p \in [2,4]$, $d_{_\text{SR}}$ represents the distance between the satellite and the UAV-relay, and $\lambda$ denotes the wavelength of the operation frequency. 

Accordingly, the received SNR of the S-R link, denoted by $\gamma_{_\text{SR}}$, can be written as follows:
\begin{equation}
	\gamma_{_\text{SR}}=\frac{P_s \left|h_{_\text{SR}}\right|^{2}}{\sigma^2_{_\text{SR}} PL_{_\text{SR}}}={\bar{\gamma}_{_\text{SR}}\left|h_{_\text{SR}}\right|^{2}}, \label{snr_sr}
\end{equation}
where $ P_s $ denotes the transmit power of the satellite, $\sigma^2_{_\text{SR}}$ denotes the variance of additive white Gaussian noise (AWGN) of the S-R link, and $\bar{\gamma}_\text{SR}$ is the average transmit SNR of the S-R link with $\bar{\gamma}_{_\text{SR}}=P_s/\left(\sigma^2_{_\text{SR}} PL_{_\text{SR}}\right)$.


\subsection{UAV relay-UE (R-U) link}
For the UAV relay-UE link, the effects of severe multi-path fading and shadowing under the post-disaster environment necessitate more appropriate channel modeling. We amend the modified-Fisher-Snedecor $\mathcal{F}$ fading channel model\cite{tcomF}, which accurately characterizes real channels under complex conditions through its dual shape parameters ($m$, $m_s$). The PDF of the channel between the UAV-relay and UE, denoted by $f_{\left|h_{_\text{RU}}\right|}(x)$, can be derived as follows:
\begin{equation}\label{hPDF}
	f_{\left|h_{_\text{RU}}\right|}(x)=\frac{2(m\Omega_s)^m(m_s\Omega_m)^{m_s}x^{2m-1}}{B(m,m_s)[m\Omega_s x^2+m_s\Omega_m]^{m+m_s}}, 
\end{equation}
where $B(a, b)$ is the beta function with $B(a,b) =\frac{\Gamma(a)\Gamma(b)}{\Gamma(a+b)}$\cite[Eq.~(8.384.1)]{Integrals} and $\Gamma(b) = \int_{0}^{\infty}{t^{b-1}e^{-t}}dt$ is the gamma function \cite[ Eq.~(8.310)]{Integrals}. $\Omega_{s}$ denotes the RMS power affected by shadowing and $\Omega_m$ represents the mean power of the multi-path component. The mean power of the composite signal envelope can be expressed as $\Omega=\Omega_m\Omega_s$ in this case. Let $t={{\left|h_{_\text{RU}}\right|}^{2}\Omega_s}/{\Omega_m}$, $t$ follows a $\mathcal{F}$ distribution with degrees of freedom $2m$ and $2m_s$ after performing variable changes and algebraic calculations. As $m\rightarrow 0, m_s\rightarrow 1$, the channel undergoes heavy multi-path fading and shadowing
\footnote{Following the derivation in  \cite{tcomF}, the mean power of the signal can be calculated by $\int_{0}^{\infty}f_{\left|h_{\tiny{_\text{RU}}}\right|}(x)\left|h_{_\text{RU}}\right|^2 d\left|h_{_\text{RU}}\right|$, which resolves to  $\mathbb{E}[{\left|h_{_\text{RU}}\right|}^2]=\frac{m_s\Omega_m}{(m_s-1)\Omega_s}=\frac{m_s\Omega}{(m_s-1)\Omega_s^2}=\Omega$. This necessitates the parameter constraint  $\Omega_{s}^2={\frac{m_s}{m_s-1}}$ to ensure non-negative power normalization. $m_s>1$ must therefore hold to maintain physical consistency in the model.}.
When $m_s \rightarrow \infty$, the shadowing eventually disappears, indicating that the Fisher-Snedecor $ \mathcal{F} $ fading reduces to Nakagami-$m$ fading\cite{Fchannel,Fchannel3}. 

We adopt the path loss model based on the statistical parameters of the underlying terrestrial environment\cite{LAP}. The path loss of the R-U link, denoted by $PL_{_\text{RU}}$ (measured in dB), is modeled as probabilistic LoS and non-line-of-sight (NLoS) links as follows:
\begin{equation}{\label{pl1}}
	PL_{_\text{RU}}=PL^{\text{dB}}_{_\text{LoS}}P_{_\text{LoS}}(\theta)+PL^{\text{dB}}_{_\text{NLoS}}P_{_\text{NLoS}}(\theta).
\end{equation}
The probability of LoS propagation, denoted by $P_{_\text{LoS}}(\theta)$, can be expressed as 
$P_{_\text{LoS}}(\theta)={1}/{\left(1+ae^{-b\theta+ab}\right)}$, 
where $\theta$ is the elevation angle. The parameters $a$ and $b$, called the S-curve parameters, are specific to the environment being simulated in \cite[Tables I-II]{LAP}. Correspondingly, the probability of NLoS propagation can be expressed as $P_{_\text{NLoS}}(\theta)=1-P_{_\text{LoS}}(\theta)$. The mean path loss of the two propagation groups is composed of free space path loss (FSPL) and excessive path loss, which can be modeled as 
\begin{equation}
	PL^{\text{dB}}_{\chi}=20\lg \frac{4\pi d_u}{\lambda}+\eta_{\chi}, \chi \in \left\{\text{LoS, NLoS}\right\}.
\end{equation}
The term $d_{u}$ is the distance between UAV relay and trapped users. 
The term $\eta_{{\chi}}$ represents excessive path loss, which largely depends on the propagation group rather than the elevation angle. 

Accordingly, the received SNR of the R-U link, denoted by $\gamma_{_\text{RU}}$, can be expressed as follows:
\begin{equation}
	\gamma_{_\text{RU}}=\frac{P_r \left|h_{_\text{RU}}\right|^{2}}{\sigma^2_{_\text{RU}} PL_{_\text{RU}}}={\bar{\gamma}_{_\text{RU}}\left|h_{_\text{RU}}\right|^{2}},
\end{equation}
where $P_r$ denotes the transmit power of the UAV-relay, $\sigma^2_{_\text{RU}}$ denotes the variance of AWGN, and $\bar{\gamma}_{_\text{RU}}$ denotes the average transmit SNR of the R-U link.


{For different relay schemes, DF relays fully decode the signal that went through the first hop and retransmit the decoded version into the second hop, offering superior performance at the cost of higher complexity and processing delay. In contrast, AF relays simply amplify and forward the incoming signal without decoding, resulting in lower complexity and latency but potentially amplifying noise alongside the desired signal. }
It is noteworthy that AF relays can be further categorized as either CSI-assisted (variable-gain) or `blind' (fixed-gain)\cite{fixedgainAF}. CSI-assisted AF relays dynamically adjust amplification using instantaneous first-hop channel state information (CSI), optimizing performance but requiring timely CSI estimation which increases complexity and overhead.  Conversely, fixed-gain AF relays employ a fixed amplification factor with no need for instantaneous CSI and thus significantly simplifying implementation and operation, albeit potentially at a performance cost. 
Furthermore, 
the variable-gain AF relaying scheme exhibits performance characteristics in its SNR expression that are often comparable to DF relaying\cite{Variablegain_AF_DF}. Therefore, focusing on the fixed-gain AF relay alongside the DF relay provides a more insightful comparison of fundamentally different relaying paradigms under the tradeoff of complexity and performance. 

The expressions of the end-to-end SNR of the composite channel under fixed-gain AF and DF relaying schemes, denoted by $\gamma_{_\text{AF}}$ and $\gamma_{_\text{DF}}$, respectively, can be expressed as 
\begin{equation}
	\begin{cases}
		\gamma_{_\text{AF}}=\dfrac{\gamma_{_\text{SR}}\gamma_{_\text{RU}}}{\left(\gamma_{_\text{RU}} + C\right)};\\
		\gamma_{_\text{DF}}=\min \left\{\gamma_{_\text{SR}},\gamma_{_\text{RU}}\right\},  
	\end{cases}
\end{equation}
respectively, where $C$ represents a parameter associated with the fixed relay gain \cite{fixedgainAF}, i.e., $C = P_r/(G^2 \sigma^2_{_\text{SR}})$ 
with $G = \sqrt{{P_r}/( {P_s\left|h_{_\text{SR}}\right|^{2} + \sigma_{_\text{SR}}^2})}$ denotes the fixed relay gain at the relay. $\min\left\{\cdot\right\} $ denotes the minimum value of the set.

\section{Statistics of SNR}\label{sec3}
In this section, we characterize the statistical properties
of the received SNR for different communication links. General analytical expressions for the PDF and CDF of the SNR for the S-R link, R-U link, fixed-gain AF relay, and DF relay are derived in the following.

\subsection{S-R link and R-U link}
\subsubsection{S-R link}
Based on Eq.~\eqref{snr_sr}, the PDF of $\gamma_{_\text{SR}}$, denoted by $f_{\gamma_{_\text{SR}}}(\gamma)$, can be expressed after some variables changes as follows:
\begin{equation}
	\begin{aligned}
		f_{\gamma_{_\text{SR}}}(\gamma) &= \frac{1}{\bar{\gamma}_{_\text{SR}}} f_{\left\lvert h_{_\text{SR}}\right\rvert ^2}\left(\frac{\gamma}{\bar{\gamma}_{_\text{SR}}}\right) \\&=  \frac{\alpha_{_\text{SR}}}{\bar{\gamma}_{_\text{SR}}} \exp\left(\frac{-(\beta_{_\text{SR}}-\delta_{_\text{SR}}) \gamma}{\bar{\gamma}_{_\text{SR}}}\right) \sum_{k=0}^{\Gamma_{s}-1} c_k\left(\frac{\gamma}{\bar{\gamma}_{_\text{SR}}}\right)^{k},
	\end{aligned}\label{PDF-SR}
\end{equation}
where we define $c_k = \frac{(-1)^k (1-\Gamma_{s})_{(k)} \delta^k_{_\text{SR}} }{(k!)^2}$\vspace{3pt} for simplicity, which is a coefficient related to $k$. Then, with the definition of the lower incomplete Gamma function $\varUpsilon(a,b)$\cite[Eq. (8.350.1)]{2007Table}, the cumulative probability function (CDF) of $\gamma_{_\text{SR}}$, denoted by $F_{\gamma_{_\text{SR}}}(\gamma)$, can be expressed as follows:
\begin{equation}
	\begin{aligned}
		F_{\gamma_{_\text{SR}}}(\gamma)=\frac{\alpha_{_\text{SR}}}{\bar{\gamma}_{_\text{SR}}}\sum_{k=0}^{\Gamma_{s}-1}\frac{c_k}{(\beta_{_\text{SR}}-\delta_{_\text{SR}})^{k+1}}\varUpsilon(k+1,\frac{(\beta_{_\text{SR}}-\delta_{_\text{SR}})\gamma}{\bar{\gamma}_{_\text{SR}}}). \label{CDF-SR}
	\end{aligned}
\end{equation}
\begin{proposition}
When $\bar{\gamma}_{_\text{SR}} \rightarrow \infty$, the exponent item-\\ $\exp\left(-\frac{(\beta_{_\text{SR}} - \delta_{_\text{SR}}) \gamma}{\bar{\gamma}_{_\text{SR}}}\right)\rightarrow 1$. Hence, the asymptotic expression of the CDF with respect to S-R link, denoted by $F_{\gamma_{_\text{SR}}}^{\text{asy}}(\gamma)$, can be expressed as follows:
\end{proposition}
\begin{equation}
	\begin{aligned}
		F_{\gamma_{_\text{SR}}}^{\text{asy}}(\gamma) &= \int_{0}^{\gamma} {\alpha_{_\text{SR}}}  \sum_{k=0}^{\Gamma_{s}-1} \frac{c_k \gamma^{k}}{ {\bar{\gamma}_{_\text{SR}}}^{k+1}} dx = \alpha_{_\text{SR}} \sum_{k=0}^{\Gamma_{s}-1} \frac{c_k}{k+1 }\left(\frac{\gamma}{\bar{\gamma}_{_\text{SR}}}\right)^{k+1}.\\
	\end{aligned}\label{ASY-CDF-SR}
\end{equation}

\subsubsection{R-U link}
Based on the distribution of $h_{_\text{RU}}$ in Eq.~\eqref{hPDF}, the PDF of $\gamma_{_\text{RU}}$, denoted by $f_{\gamma_{_\text{RU}}}(\gamma)$, can be expressed as follows:
\begin{equation}
	f_{\gamma_{_\text{RU}}}(\gamma)=\frac{(m\Omega_s)^m(m_s\Omega_m \bar{\gamma}_{_\text{RU}})^{m_s}{{\gamma}}^{m-1}}{B(m,m_s)\left(m\Omega_s{\gamma}+m_s\Omega_m\bar{\gamma}_{_\text{RU}}\right) ^{m+m_s}}.\label{PDF-RU}
\end{equation}
Correspondingly, the CDF expression of $\gamma_\text{RU}$, denoted by $F_{\gamma_{_\text{RU}}}(\gamma)$, can be expressed as follows:
\begin{equation}
	\begin{aligned}
		&F_{\gamma_{_\text{RU}}}(\gamma)=F_{h_{_\text{RU}}}(\sqrt{\frac{\gamma}{\bar{\gamma}_{_\text{RU}}}})\\
		&\overset{(a)}{=} \left(\frac{m\Omega_{s}\gamma}{m_s\Omega_m\bar{\gamma}_{_\text{RU}}}\right)^m\frac{{}_2 F_1\left(m+m_s,m;m+1;-\frac{m\Omega_{s}\gamma}{m_s\Omega_m\bar{\gamma}_{_\text{RU}}}\right)}{B(m,m_s)m}\\
		&\overset{(b)}{=} \frac{1}{\Gamma(m)\Gamma(m_s)} G_{2,2}^{2,1}\left( \frac{m_s\Omega_m\bar{\gamma}_{_\text{RU}}}{m\Omega_s\gamma} \Bigg| \begin{gathered}
			{1-m, 1}\\{m_s, 0}\end{gathered} \right),  \label{CDF-RU}
	\end{aligned}
\end{equation}
where ${}_2 F_1\left(\alpha, \beta; \gamma; -z\right)$ denotes the hypergeometric function\\\cite[Eq.~(9.111)]{2007Table} and $G_{p, q}^{m, n}\left( x\Bigg| \\\begin{gathered}
	{a_{1}, \cdots, a_{p}}\\{b_{1}, \cdots, b_{q}}
\end{gathered}\right)$ is the Meijer's G-function \cite[Eq.(9.301)]{2007Table}. $(a)$ results from \cite[Eq.~(3.194.1)]{2007Table},  $(b)$ results from \cite[Eq.~(9.31.5)]{2007Table}, i.e., $x^k G_{p,q}^{m,n}\left(x \Bigg|  \begin{gathered}
{\bf{a_{p}}}\\\bf{b_{q}}\end{gathered}\right)=G_{p,q}^{m,n}\left(x \Bigg|  \begin{gathered}{{\bf a_{p}}+k}\\{\bf b_{q}}+k\end{gathered}\right)$ and \cite[Eq.~(9.34.8)]{2007Table}, i.e., $ \begin{aligned}{ }_{p} F_{q}\left({\bf a_{p}};{\bf b_{q}}; x\right) & =\frac{\prod_{j=1}^{q} \Gamma\left(b_{j}\right)}{\prod_{j=1}^{p} \Gamma\left(a_{j}\right)} G_{q+1, p}^{p, 1}\left(-\frac{1}{x} \left\lvert\, \begin{array}{l}1,{\bf b_{q}}\\ {\bf a_{p}}\end{array}\right.\right)\end{aligned}$.\vspace{5pt}
\begin{proposition}
When $\bar{\gamma}_{_\text{RU}} \rightarrow \infty$, we have $\frac{m\Omega_s \gamma}{m_s\Omega_m\bar{\gamma}_\text{RU}}\rightarrow 0$\vspace{3pt}. According to \cite[Eq.~(07.23.03.0001.01)]{walfram} (i.e., ${}_2 F_1\left(\alpha, \beta; \gamma; 0\right) = 1$), the asymptotic expression of $F_{\gamma_{_\text{RU}}}(\gamma)$ can be further expressed as follows:
\end{proposition}
\begin{equation}
	F_{\gamma_{_\text{RU}}}^{\text{asy}}(\gamma)=\frac{m^{m-1}\Omega_s^m \gamma^m}{B(m,m_s)(m_s\Omega_m\bar{\gamma}_{_\text{RU}})^m}. \label{ASY-CDF-RU}
\end{equation}

\subsection{Fixed-gain AF relay and DF relay}
\subsubsection{Fixed-gain AF relay} For fixed-gain AF relaying system, the CDF expression of $\gamma_{_\text{AF}}$, denoted by $F_{\gamma_{_\text{AF}}}(\gamma)$, can be expressed as follows: 
\begin{equation}
	\begin{aligned} 
		F_{\gamma_{_\text{AF}}}(\gamma) &= \frac{\alpha_{_\text{SR}}}{\bar{\gamma}_{_\text{SR}}}\sum_{k=0}^{\Gamma_{s}-1}\frac{c_k}{(\beta_{_\text{SR}}-\delta_{_\text{SR}})^{k+1}}\varUpsilon(k+1,\frac{(\beta_{_\text{SR}}-\delta_{_\text{SR}})\gamma}{\bar{\gamma}_{_\text{SR}}})\\
		&+\Lambda \sum_{k=0}^{\Gamma_{s}-1} \frac{c_k}{{\bar{\gamma}_{_\text{SR}}}^{k+1}} \left[\sum_{i=0}^{k}  \left(\frac{\beta_{_\text{SR}} - \delta_{_\text{SR}}}{\bar{\gamma}_{_\text{SR}}}\right)^{-(i+1)} \gamma^{k-i} \right.\\
		&\times \left. G_{3,2}^{2,2}\left(\frac{m_s\Omega_m\bar{\gamma}_{_\text{SR}}\bar{\gamma}_{_\text{RU}}}{(\beta_{_\text{SR}} - \delta_{_\text{SR}})m\Omega_sC\gamma}  \mid \begin{gathered}
			{-i, 1-m, 1}\\{m_s, 0}\end{gathered} \right)\right],
		\label{CDF-AF}
	\end{aligned}
\end{equation}
where $\Lambda = \frac{\alpha_{_\text{SR}} \exp\left(-\frac{(\beta_{_\text{SR}} - \delta_{_\text{SR}}) \gamma}{\bar{\gamma}_{_\text{SR}}}\right)}{\Gamma(m)\Gamma(m_s)}$ is defined for simplicity.
\begin{IEEEproof} 
Since S-R and R-U links suffer fading and shadowing independently, the CDF expression of $\gamma_{_\text{AF}}$ can be expressed as \cite{AF-YANG,AFasy}
\begin{equation}
		\begin{aligned} 
			F_{\gamma_{_\text{AF}}}(\gamma) &=\int_{0}^{\infty}\text{Pr} \left[\dfrac{\gamma_{_\text{SR}}\gamma_{_\text{RU}}}{\gamma_{_\text{RU}}+C}<\gamma \Bigg|  \gamma_{_\text{SR}}\right] f_{\gamma_{_\text{SR}}} \left(\gamma_{_\text{SR}}\right) d \gamma_{_\text{SR}}\\ 
			&=F_{\gamma_{_\text{SR}}}(\gamma) + \underbrace{\int_{0}^{\infty} F_{\gamma_{_\text{RU}}}(\frac{C\gamma}{x}) f_{\gamma_{_\text{SR}}}(x+\gamma) dx}_{\mathcal{L} }, \label{AF-L0}
		\end{aligned}
\end{equation}
where $F_{\gamma_{_\text{SR}}}(\gamma)$ is presented in Eq.~\eqref{CDF-SR}. Substituting Eqs.~\eqref{PDF-SR} and (\ref{CDF-RU}) into Eq.~(\ref{AF-L0}), the integral $\mathcal{L}$ can be calculated as follows:
	\begin{equation}
		\begin{aligned}
			\mathcal{L} &= \frac{\alpha_{_\text{SR}}\exp\left(-\frac{(\beta_{_\text{SR}} - \delta_{_\text{SR}}) \gamma}{\bar{\gamma}_{_\text{SR}}}\right)}{\bar{\gamma}_{_\text{SR}}\Gamma(m)\Gamma(m_s)} \int_{0}^{\infty} \exp\left(-\frac{(\beta_{_\text{SR}} - \delta_{_\text{SR}}) x}{\bar{\gamma}_{_\text{SR}}}\right)\\ &\times G_{2,2}^{2,1}\left( \frac{m_s\Omega_m\bar{\gamma}_{_\text{RU}}}{m\Omega_sC\gamma}x \Bigg| \begin{gathered}
				{1-m, 1}\\{m_s, 0}\end{gathered} \right)
			\sum_{k=0}^{\Gamma_{s}-1} c_k\left(\frac{x+\gamma}{ {\bar{\gamma}_{_\text{SR}}}}\right)^kdx.
			\label{AF-L1}
		\end{aligned}
	\end{equation}
According to \cite[Eq.~({07.34.03.0228.01})]{walfram}, the exponential term \vspace{3pt}$\exp\left(-\frac{(\beta_{_\text{SR}} - \delta_{_\text{SR}}) x}{\bar{\gamma}_{_\text{SR}}}\right)$ can be transformed to  $G_{0,1}^{1,0}\left(\frac{(\beta_{_\text{SR}} - \delta_{_\text{SR}})}{\bar{\gamma}_{_\text{SR}}}x \Bigg|  0 \right) $. 
Since Eq.~(\ref{AF-L1}) is intractable for direct evaluation, we restrict our analysis to the case where $\Gamma_{s}$ is an integer, thereby enabling binomial expansion with
$(x+\gamma)^k = \sum_{r=0}^{k} \binom{k}{r} x^{k-r} \gamma^r$.
Then using \cite[Eq.~(7.811.1)]{2007Table} and after some manipulations, the integral can be calculated as follows:
	\begin{equation}
		\begin{aligned}
			\mathcal{L} &= \Lambda \sum_{k=0}^{\Gamma_{s}-1} \frac{c_k}{{\bar{\gamma}_{_\text{SR}}}^{k+1}} \left[\sum_{i=0}^{k} \binom{k}{i} \left(\frac{\beta_{_\text{SR}} - \delta_{_\text{SR}}}{\bar{\gamma}_{_\text{SR}}}\right)^{-(i+1)} \gamma^{k-i}\right.\\  &\times\left. G_{3,2}^{2,2}\left(\frac{m_s\Omega_m\bar{\gamma}_{_\text{SR}}\bar{\gamma}_{_\text{RU}}}{(\beta_{_\text{SR}} - \delta_{_\text{SR}})m\Omega_sC\gamma} \Bigg| \begin{gathered}
				{-i, 1-m, 1}\\{m_s, 0}\end{gathered} \right)\right].
			\label{AF-L3}
		\end{aligned}
	\end{equation}
By substituting Eq.~(\ref{AF-L3}) into Eq.~(\ref{AF-L0}), Eq.~(\ref{CDF-AF}) is obtained, thus completing the proof. 
\end{IEEEproof}
\begin{proposition}
	With high SNR, the asymptotic expression of the CDF for fixed-gain AF relay, denoted by $F^{\text{asy}}_{\gamma_{_\text{AF}}}(\gamma)$, can be expressed as follows:
	\begin{equation}
		\begin{aligned}
			&F^{\text{asy}}_{\gamma_{_\text{AF}}}(\gamma) = \alpha_{_\text{SR}} \sum_{k=0}^{\Gamma_{s}-1} \frac{c_k}{k+1}\left(\frac{\gamma}{\bar{\gamma}_{_\text{SR}}}\right)^{k+1} \\ &\times \left[1 + \frac{\Gamma(m-k-1)\Gamma(m_s+k+1)}{\Gamma(m_s)\Gamma(m)} \left(\frac{m\Omega_{s}C}{m_s\Omega_m\bar{\gamma}_{_\text{RU}}}\right)^{k+1}\right].\label{asy-AF-final}
		\end{aligned}
	\end{equation}\end{proposition}
\begin{IEEEproof} 
	In order to provide more concise expressions of the system performance, an asymptotic expression of the CDF for fixed-gain AF relaying can be derived as 
	\begin{equation}
		F^{\text{asy}}_{\gamma_{_\text{AF}}}(\gamma) =F^{\text{asy}}_{\gamma_{_\text{SR}}}(\gamma) + {\mathcal{L}^{\text{asy}}},\label{AF-asy1}
	\end{equation}	
	where $F^{\text{asy}}_{\gamma_{_\text{SR}}}(\gamma)$ is the asymptotic expression given in Eq.~(\ref{ASY-CDF-SR}), and $\mathcal{L}^{\text{asy}}$ denotes the asymptotic result of integral $\mathcal{L}$. In similar, when $\bar{\gamma} \rightarrow \infty$, the exponent item $\exp\left(-\frac{(\beta_{_\text{SR}} - \delta_{_\text{SR}}) (x+\gamma)}{\bar{\gamma}_{_\text{SR}}}\right)$ approaches to 1 and the term $\left(\frac{x+\gamma}{ {\bar{\gamma}_{_\text{SR}}}}\right)^k$ converges to $\left(\frac{x}{{\bar{\gamma}_{_\text{SR}}}}\right)^k$. Therefore, $\mathcal{L}^{\text{asy}}$ can be further asymptotically expressed as follows:
	\begin{equation}
		\begin{aligned}
		\mathcal{L}^{\text{asy}} &=\frac{\alpha_{_\text{SR}}}{\Gamma(m)\Gamma(m_s)}\\
		&\times \int_{0}^{\infty} G_{2,2}^{2,1}\left( \frac{m_s\Omega_m\bar{\gamma}_{_\text{RU}}}{m\Omega_{s}C\gamma}x \Bigg| \begin{gathered}
				{1-m, 1}\\{m_s, 0}\end{gathered} \right) \sum_{k=0}^{\Gamma_{s}-1} \frac{c_k x^k}{{\bar{\gamma}_{_\text{SR}}}^{k+1}} dx.
		\end{aligned}
	\end{equation}
	Then utilizing \cite[Eq.~(7.811.4)]{2007Table}, we obtain 
	\begin{equation}
		\begin{aligned}
		\mathcal{L}^{\text{asy}} =\alpha_{_\text{SR}} \sum_{k=0}^{\Gamma_{s}-1}&\left[\frac{\Gamma(m-k-1)\Gamma(m_s+k+1) c_k}{\Gamma(m_s)\Gamma(m)(k+1)}\right.\\
			&\times \left.\left(\frac{m\Omega_{s}C\gamma}{m_s\Omega_m\bar{\gamma}_{_\text{SR}}\bar{\gamma}_{_\text{RU}}}\right)^{k+1}\right].\label{L-asy}		
			\end{aligned}
	\end{equation}
	By substituting Eq.~(\ref{L-asy}) into Eq.~(\ref{AF-asy1}), Eq.~(\ref{asy-AF-final}) can be deduced, thereby completing the proof.    
	\end{IEEEproof}     

\subsubsection{DF relay}: for DF relaying system, the CDF of $\gamma_{_\text{DF}}$, denoted by $F_{\gamma_{_\text{DF}}}(\gamma)$, can be expressed as follows\cite{DF}:
\begin{equation}
	\begin{aligned} F_{\gamma_{_\text{DF}}}(\gamma)  &=  \text{Pr}\left(\min \left\{\gamma_{_\text{SR}},\gamma_{_\text{RU}}\right\} \le \gamma\right) \\
		&= 1-\left(1-F_{\gamma_{_\text{SR}}}(\gamma) \right)\left(1-F_{\gamma_{_\text{RU}}}(\gamma) \right)\\
		&=F_{\gamma_{_\text{SR}}}(\gamma)+F_{\gamma_{_\text{RU}}}(\gamma)-F_{\gamma_{_\text{SR}}}(\gamma)F_{\gamma_{_\text{RU}}}(\gamma).\label{CDF-DF}
	\end{aligned}
\end{equation}
Substituting Eqs.~\eqref{CDF-SR} and \eqref{CDF-RU} into Eq.~\eqref{CDF-DF} yields the exact CDF expression of $\gamma_{_\text{DF}}$.

Then, the PDF expression can be obtained by taking the derivative of CDF with respect to $\gamma$ as follows:
\begin{equation}
	f_{\gamma_{_\text{DF}}}(\gamma)=f_{\gamma_{_\text{SR}}}(\gamma)+f_{\gamma_{_\text{RU}}}(\gamma)-f_{\gamma_{_\text{SR}}}(\gamma) F_{\gamma_{_\text{RU}}}(\gamma)-f_{\gamma_{_\text{RU}}}(\gamma) F_{\gamma_{_\text{SR}}}(\gamma)\label{PDF-DF}.
\end{equation}
\begin{proposition}
With high SNR, the asymptotic CDF for DF relay simplifies to\end{proposition}
\begin{equation}
	\begin{aligned}
		F^{\text{asy}}_{\gamma_{_\text{DF}}}(\gamma) &=F^{\text{asy}}_{\gamma_{_\text{SR}}}(\gamma)+F^{\text{asy}}_{\gamma_{_\text{RU}}}(\gamma) - F^{\text{asy}}_{\gamma_{_\text{SR}}}(\gamma)F^{\text{asy}}_{\gamma_{_\text{RU}}}(\gamma)\\
		&\approx F^{\text{asy}}_{\gamma_{_\text{SR}}}(\gamma)+F^{\text{asy}}_{\gamma_{_\text{RU}}}(\gamma).
	\end{aligned}\label{asy-DF-final}
\end{equation}

\section{Performance Analysis}\label{sec4}

In this section, we derive theoretical expressions for the effective capacity, the outage probability, and the $\epsilon${-}outage capacity of the considered system. 
Additionally, we provide an asymptotic analysis of the outage probability in high SNR regimes, based on which the corresponding diversity orders are obtained. 
These results serve to guide the pre-deployment assessment of EWC systems under stringent latency and reliability requirements, help determine coverage boundaries, and inform design trade-offs in power control and relay selection strategies.

\subsection{Effective Capacity}
Considering different QoS requirements for various services under EWC scenarios, the statistical delay-bounded QoS exponent is used to quantify the delay-QoS requirements. 
Based on large deviation principle (LDP), under sufficient conditions, the queue length process $Q(t)$ converges in distribution to a random variable $Q(\infty)$ such that
\begin{equation}
	-\lim _{x \rightarrow \infty} \frac{\log (\operatorname{Pr}\{Q(\infty)>Q(\text{th})\})}{Q(\text{th})}=\theta, 
\end{equation}
where $Q(\text{th})$ denotes the queue length bound. The QoS exponent, denoted by $\theta~(\theta>0)$, represents the exponential decay rate of delay-bound QoS violation probabilities.  A larger $\theta$ (a faster decay rate) indicates a stringent QoS requirement, while a smaller $\theta$ (a smaller decay rate) implies a more relaxed QoS standard. Particularly, when $\theta \rightarrow 0$, arbitrary long delay can be tolerated. Conversely, if $\theta \rightarrow \infty$, the system cannot tolerate any delay.

According to \cite{2007qos}, the effective
capacity of the service process for a given $\theta$, denoted by $\mathrm{E}_{\mathrm{C}}(\theta)$,  can be expressed as follows:
\begin{equation}
	\begin{aligned}
		\mathrm{E}_{\mathrm{C}}(\theta)& = -\frac{1}{\theta}\log \left(\mathbb{E}_\gamma\left[e^{-R\theta}\right]\right)= -\frac{1}{\theta}\log \left(\mathbb{E}_\gamma\left[\left(1+\gamma\right)^{-\beta} \right]\right)\\
		&=-\frac{1}{\theta} \log \left(\int_{0}^{\infty} \frac{f_{\gamma}(\gamma)}{\left(1+{\gamma}\right)^{\beta}} d \gamma\right),
	\end{aligned}
	\label{EC}  
\end{equation}
where $R$ denotes the instantaneous service rate, i.e., Shannon capacity, defined as $R=BT\log(1+\gamma)$. $B$ is the system spectral bandwidth, $\mathbb{E}_\gamma\left[\cdot\right] $ represents the expectation of $\gamma$, $T$ denotes the time duration, and $\beta=\frac{BT\theta}{\log2}$ represents the normalized QoS exponent. Since the expressions of $f_{\gamma}(\gamma)$ are presented, the effective capacity can be obtained in the following. 

\subsubsection{S-R link}
With the help of \cite[Eq.~(07.34.03.0271.01)]{walfram}, $\dfrac{1}{(1+\gamma)^\beta}$ can be rewritten as $\dfrac{1}{\Gamma(\beta)} G^{1,1}_{1,1}\left(\gamma \middle|\, \begin{gathered}
	1-\beta\\
	0
\end{gathered}\right)$. Substituting Eq.~(\ref{PDF-SR}) into Eq.~(\ref{EC}), applying \cite[Eq.~(9.31.5)]{2007Table}, the effective capacity of the S-R link, denoted by $\mathrm{E}_{\mathrm{C}}^{\text{SR}}(\theta)$, can be reformulated as 
\begin{equation}
	\begin{aligned}
		\mathrm{E}_{\mathrm{C}}^{\text{SR}}(\theta)&= -\frac{1}{\theta} \log \left[\frac{\alpha_{_\text{SR}}}{\Gamma(\beta)}\sum_{k=0}^{\Gamma_{s}-1} \frac{c_k \gamma^k}{{\bar{\gamma}_{_\text{SR}}}^{k+1}} 
     		\right.\\
		&\times \left.
				G^{1,1}_{1,1}\left(\gamma \middle|\, \begin{gathered}
			1-\beta\\
			0
		\end{gathered}\right) G^{1,0}_{0,1}\left(\frac{\left(\beta_{_\text{SR}}-\delta_{_\text{SR}}\right)\gamma}{\bar{\gamma}_{_\text{SR}}} \middle|\, 0\right) \right].\label{EC-SR1}
	\end{aligned}
\end{equation}
With the use of \cite[Eq.~(7.811.1)]{2007Table}, $\mathrm{E}_{\mathrm{C}}^{\text{SR}}(\theta)$ can be derived as follows:
\begin{equation}
	\begin{aligned}
		\mathrm{E}_{\mathrm{C}}^{\text{SR}}(\theta)&= -\frac{1}{\theta} \log \left[\frac{\alpha_{_\text{SR}}}{\Gamma(\beta)}\sum_{k=0}^{\Gamma_{s}-1} \frac{c_k}{{\bar{\gamma}_{_\text{SR}}}^{k+1}}
		\right.\\
		&\quad \times \left.
		G^{2,1}_{1,2}\left(\frac{\beta_{_\text{SR}}-\delta_{_\text{SR}}}{\bar{\gamma}_{_\text{SR}}} \middle|\, \begin{gathered}
			-k\\
			\beta-k-1, 0
		\end{gathered}\right) \right].\label{EC-SR2}
	\end{aligned}
\end{equation}

\subsubsection{R-U link}
According to the properties of the hypergeometric function and Meiger-G function \cite[Eqs.~({07.19.02.0002.01}) and ({07.34.03.0006.01})]{walfram}, the PDF of $\gamma_{_\text{RU}}$ can be rewritten as
\begin{equation}
	f_{\gamma_{_\text{RU}}}(\gamma)=	\frac{m\Omega_s}{\Gamma(m)\Gamma(m_s)m_s\Omega_m \bar{\gamma}_{_\text{RU}}} G^{1,1}_{1,1}\left({\frac{m\Omega_s\gamma}{m_s\Omega_m \bar{\gamma}_{_\text{RU}}}} \middle|\, \begin{gathered}
		-m_s\\
		m-1
	\end{gathered}\right) .\label{PDF-RU2}
 \end{equation}
Similarly, substituting Eq.~(\ref{PDF-RU2}) into Eq.~(\ref{EC}) and applying \cite[Eq.~(07.34.03.0271.01)]{walfram}, the effective capacity of the R-U link, denoted by $\mathrm{E}_{\mathrm{C}}^{\text{RU}}(\theta)$, can be deduced as follows:
\begin{equation}
	\begin{aligned}
		\mathrm{E}_{\mathrm{C}}^{\text{RU}}(\theta)
		& =-\frac{1}{\theta} \log \left[ \int_{0}^{\infty}
		\frac{m\Omega_s}{\Gamma(\beta)\Gamma(m)\Gamma(m_s)m_s\Omega_m \bar{\gamma}_{_\text{RU}}} \right.\\
		&\times\left.G^{1,1}_{1,1}\left(\gamma \middle|\, \begin{gathered}
			1-\beta\\
			0
		\end{gathered}\right)
		G^{1,1}_{1,1}\left({\frac{m\Omega_s\gamma}{m_s\Omega_m \bar{\gamma}_{_\text{RU}}}} \middle|\, \begin{gathered}
			-m_s\\
			m-1
		\end{gathered}\right) d\gamma \right].
	\end{aligned}
\end{equation}
Then using \cite[Eqs.~(7.811.1) and (9.31.5)]{2007Table}, $\mathrm{E}_{\mathrm{C}}^{\text{RU}}(\theta)$ can be obtained as follows:
\begin{equation}
	\begin{aligned}
		\mathrm{E}_{\mathrm{C}}^{\text{RU}}(\theta)
		=-\frac{1}{\theta} \log & \left[
		\frac{1}{\Gamma(m)\Gamma(m_s)\Gamma(\beta)} \right.\\
		&\times\left.
		G^{2,2}_{2,2}\left(\frac{m\Omega_s}{m_s\Omega_m\bar{\gamma}_{_\text{RU}}} \middle|\, \begin{gathered}
			1, 1-m_s\\
			\beta, m
		\end{gathered}\right)\right].\label{EC-RU}
	\end{aligned}
\end{equation}

\subsubsection{Fixed-gain AF relay}
Based on Eqs.~(\ref{AF-L0}) and (\ref{EC}), the effective capacity of the fixed-gain AF relay, denoted by $\mathrm{E}_{\mathrm{C}}^{\text{AF}}(\theta)$, 
can be expressed after performing variable changes as follows:
\begin{equation}
	\begin{aligned}
		\mathrm{E}_{\mathrm{C}}^{\text{AF}}(\theta)&=-\frac{1}{\theta} \log \left(\beta \int_{0}^{\infty} \frac{F_{\gamma_{_\text{AF}}}(\gamma)}{\left(1+{\gamma}\right)^{\beta+1}} d \gamma\right)\\
		&=-\frac{1}{\theta} \log \left(\beta \underbrace{\int_{0}^{\infty} \frac{F_{\gamma_{_\text{SR}}}(\gamma)}{\left(1+{\gamma}\right)^{\beta+1}}}_{\mathcal{M}_1}+\underbrace{\frac{\mathcal{L}}{\left(1+{\gamma}\right)^{\beta+1}} d \gamma}_{\mathcal{M}_2} \right).\label{EC-AF} 
	\end{aligned}
\end{equation}
According to \cite[Eq.~(8.356.1)]{2007Table}, \vspace{3pt}i.e., $\Gamma(a, z)+\varUpsilon(a, z)=\Gamma(a)$, and \cite[Eq.~(06.06.26.0005.01)]{walfram}, i.e., $ \Gamma(a, z)=G_{1,2}^{2,0}\\\left(z\Bigg| \begin{array}{l} 1\\ 0, a\end{array}\right)$, the lower incomplete gamma function in $F_{\gamma_{_\text{SR}}}(\gamma)$ can be rewritten as
\begin{equation}
	\begin{aligned}
		\varUpsilon(k+1,\frac{(\beta_{_\text{SR}}-\delta_{_\text{SR}})\gamma}{\bar{\gamma}_\text{SR}})&=\Gamma(k+1)\\ &-G_{1,2}^{2,0}\left(\frac{(\beta_{_\text{SR}}-\delta_{_\text{SR}})\gamma}{\bar{\gamma}_{_\text{SR}}} \Bigg| \begin{array}{c}1 \\ 0, k+1\end{array}\right),\label{lower-gamma}    
	\end{aligned}
\end{equation}
where $\Gamma(a, z)$ is the upper incomplete gamma function\cite[Eq.~(8.350.2)]{2007Table}. By employing  Eq.~\eqref{lower-gamma} and \cite[Eq.~(7.811.5)]{2007Table}, after numerical integral, $\mathcal{M}_1$ can be deduced as follows:
\begin{equation}
	\begin{aligned}
		\mathcal{M}_1&=\sum_{k=0}^{\Gamma_{s}-1} \frac{\alpha_{_\text{SR}}c_k }{\bar{\gamma}_{_\text{SR}}(\beta_{_\text{SR}}-\delta_{_\text{SR}})^{k+1}} \\
		&\times \Bigg[\frac{\Gamma(k+1)}{\beta}-\frac{G_{2,3}^{3,1}\left(\frac{\beta_{_\text{SR}}-\delta_{_\text{SR}}}{\bar{\gamma}_\text{SR}} \Bigg| \begin{array}{l}0, 1 \\\beta, 0, k+1\end{array}\right)}{\Gamma(\beta+1)}\Bigg].\label{EC-AF-M1} 
	\end{aligned}
\end{equation}

Based on the reciprocal property of Meijer G-functions \cite[Eq.~(9.31.2)]{2007Table}, i.e., $ G_{p,q}^{m,n}\left(\begin{array}{l|l}x^{-1} & \begin{array}{l}a_{r} \\ b_{s}\end{array}\end{array}\right)=G_{q,p}^{n,m}\left(x \left\lvert\, \begin{array}{l}1-b_{s} \\ 1-a_{r}\end{array}\right.\right)$,\vspace{3pt} applying \cite[Eqs.~(7.811.5), (9.31.5)]{2007Table} and \cite[Eq.~({07.34.03.0228.01})]{walfram}, Eq.~\eqref{AF-L3} can be rewritten as follows:
\begin{equation}
	\begin{aligned}
		\mathcal{L} = \sum_{k=0}^{\Gamma_{s}-1} d_k &\sum_{i=0}^k \binom{k}{i}\left[ G_{0,1}^{1,0}\left(\frac{(\beta_{_\text{SR}} - \delta_{_\text{SR}})}{\bar{\gamma}_{_\text{SR}}} \gamma \Bigg| k-i \right)\right. \\ &  \times\left.G_{2,3}^{2,2}\left(\frac{(\beta_{_\text{SR}} - \delta_{_\text{SR}})m\Omega_{s}C\gamma}{m_s\Omega_m\bar{\gamma}_{_\text{SR}}\bar{\gamma}_{_\text{RU}}}  \Bigg| \begin{gathered}
			{1-m_s, 1}\\{m, 1+i, 0}\end{gathered} \right)\right].  \label{AF-L4} 
	\end{aligned}
\end{equation}
where $d_k = \frac{\alpha_{_\text{SR}}c_k}{\Gamma(m)\Gamma(m_s)(\beta_{_\text{SR}}-\delta_{_\text{SR}})^{k+1}}$.\vspace{3pt}
Then applying Eq.~\eqref{AF-L4} and \cite[Eqs.~(07.34.03.0271.01)]{walfram}, the integral $\mathcal{M}_2$ can be deduced as Eq.~\eqref{EC-AF-M222} shown at the top of this page, where $(a)$ results from \cite [Eq.~(07.34.21.0081.01)]{walfram}, $G^{m,n:s_1,t_1:s_2,t_2}_{p,q:u_1,v_1:u_2,v_2}
\left(\begin{array}{c|c|c|} \bf{a_p} & \bf{c_{u_1}} & \bf{c_{u_2}}\\ \bf{b_q} &\bf{d_{v_1}} & \bf{d_{v_2}}\\ \end{array} z,w\right)$ denotes the Bivariate Meijer G-Function.
Subsequently, the closed-form expression of $\mathrm{E}_{\mathrm{C}}^{\text{AF}}(\theta)$ can be obtained by substituting the results of $\mathcal{M}_1$ and $\mathcal{M}_2$ into Eq.~(\ref{EC-AF}).

\begin{figure*}
	\begin{equation}
		\begin{aligned}
			\mathcal{M}_2 &=\int_{0}^{\infty} \sum_{k=0}^{\Gamma_{s}-1} \frac{d_k}{\Gamma(\beta)} G_{1,1}^{1,1}\left(\gamma \Bigg| \begin{array}{l}1-\beta\\0\end{array}\right)  \sum_{i=0}^k  \binom{k}{i} G_{0,1}^{1,0}\left(\frac{(\beta_{_\text{SR}} - \delta_{_\text{SR}})}{\bar{\gamma}_{_\text{SR}}} \gamma \Bigg| k-i \right) G_{2,3}^{2,2}\left(\frac{(\beta_{_\text{SR}} - \delta_{_\text{SR}})m\Omega_{s}C\gamma}{m_s\Omega_m\bar{\gamma}_{_\text{SR}}\bar{\gamma}_{_\text{RU}}}  \Bigg| \begin{gathered}
				{1-m_s, 1}\\{m, 1+i, 0}\end{gathered} \right) d\gamma \\
			&\overset{(a)}{=} \sum_{k=0}^{\Gamma_{s}-1} \frac{d_k}{\Gamma(\beta)} \sum_{i=0}^k \binom{k}{i} G^{1,1:1,0:2,2}_{1,1:0,1:2,3}
			\left(\begin{array}{c|c|ccc|} 0 & &1-m_s,&1&\\ \beta-1 &k-i&m,&1+i,&0\\ \end{array} \frac{(\beta_{_\text{SR}} - \delta_{_\text{SR}})}{\bar{\gamma}_{_\text{SR}}},\frac{(\beta_{_\text{SR}} - \delta_{_\text{SR}})m\Omega_sC}{\bar{\gamma}_{_\text{SR}}\bar{\gamma}_{_\text{RU}} m_s\Omega_m}\right), \label{EC-AF-M222} 
		\end{aligned}
	\end{equation} 
	\rule[-7pt]{18.05cm}{0.05em}   
\end{figure*}
\begin{proposition}
The asymptotic expression for the effective capacity of   fixed-gain AF relay, denoted by $\widetilde{\mathrm{E}}_{\mathrm{C}}^{\text{AF}}(\theta)$, can be expressed as follows:
\begin{equation}
	\begin{aligned}
	&\widetilde{\mathrm{E}}_{\mathrm{C}}^{\text{AF}}(\theta) = -\frac{1}{\theta} \log \left(\beta \sum_{k=0}^{\Gamma_{s}-1} e_k B(k+2, \beta -k-1) \right),\label{asy-EC-AF}
	\end{aligned}
\end{equation}\end{proposition}
where 
\begin{equation}
	\begin{aligned}
	e_k &= \frac{\alpha_{_\text{SR}} c_k}{(k+1){\bar{\gamma}_{_\text{SR}}}^{k+1}} \\&\times\left[1 +\frac{\Gamma(m-k-1)\Gamma(m_s+k+1)}{\Gamma(m_s)\Gamma(m)}
	 \left(\frac{m\Omega_{s}C}{m_s\Omega_m\bar{\gamma}_{_\text{RU}}}\right)^{k+1}\right].
	\end{aligned}
\end{equation}
\begin{IEEEproof} 
The expression of $
\mathcal{M}_2$ in Eq.~\eqref{EC-AF-M222} is intractable for subsequent computation and optimization due to the complex Meijer-G function involved, hence the derived asymptotic expression in Section \ref{sec3} is employed to present asymptotic results. Plugging Eq.~\eqref{asy-AF-final} into Eq.~\eqref{EC-AF}, we have 
\begin{equation}
	\begin{aligned}
	&\widetilde{\mathrm{E}}_{\mathrm{C}}^{\text{AF}}(\theta) = -\frac{1}{\theta} \log \left(\beta {\int_{0}^{\infty} \sum_{k=0}^{\Gamma_{s}-1}}\frac{e_k \gamma^{k+1}}{(1+\gamma)^{\beta+1}}d\gamma\right).
	\end{aligned}
\end{equation}
Consequently, applying \cite[Eq.~(3.194.3)]{2007Table}, the asymptotic effective capacity for fixed-AF relay is obtained as Eq.~\eqref{asy-EC-AF}, which completes the proof. 
\end{IEEEproof} 
\subsubsection{DF relay}
Substituting Eq.~\eqref{PDF-DF} into Eq.~(\ref{EC}), the effective capacity of the DF relay, denoted by $\mathrm{E}_{\mathrm{C}}^{\text{DF}}(\theta)$, can be expressed as follows:
\begin{equation}
	\begin{aligned}
		\mathrm{E}_{\mathrm{C}}^{\text{DF}}(\theta)& = -\frac{1}{\theta} \log \Big [\underbrace{ \int_{0}^{\infty}\frac{f_{\gamma_{_\text{SR}}}(\gamma)+f_{\gamma_{_\text{RU}}}(\gamma)}{(1+\gamma)^{\beta}}}_{\mathcal{N}}\\
		&-\underbrace{\frac{f_{\gamma_{_\text{SR}}}(\gamma) F_{\gamma_{_\text{RU}}}(\gamma)}{(1+\gamma)^{\beta}}}_{\mathcal{N}_1}-\underbrace{\frac{f_{\gamma_{_\text{RU}}}(\gamma) F_{\gamma_{_\text{SR}}}(\gamma)}{(1+\gamma)^{\beta}}d\gamma}_{\mathcal{N}_2} \Big ],\label{EC-DF}
	\end{aligned}
\end{equation}
where the closed-form expression of $\mathcal{N}$ can be obtained directly by adding the results of Eqs.~(\ref{EC-SR2}) and (\ref{EC-RU}).

By substituting $f_{\gamma_{_\text{SR}}}(\gamma) $ and $F_{\gamma_{_\text{RU}}}(\gamma)$ in Eqs.\eqref{PDF-SR} and \eqref{CDF-RU}, respectively, the closed-form expression for ${\mathcal{N}_1}$ can be obtained by calculating the integral of Meijer-G function as shown in Eq.~\eqref{EC-AF-N1} at the top of next page, where $(a)$ results from \cite[Eq.~(07.34.03.0271.01)]{walfram} and \cite[Eq.~({07.34.03.0228.01})]{walfram}, $(b)$ results from \cite [Eq.~(07.34.21.0081.01)]{walfram}.
\begin{figure*}
	\begin{equation}
		\begin{aligned}
		\mathcal{N}_1 &\overset{(a)}{=}\int_{0}^{\infty}\frac{\alpha_{_\text{SR}}}{\Gamma(m)\Gamma(m_s)\Gamma(\beta)} \sum_{k=0}^{\Gamma_{s}-1} \frac{c_k}{\bar{\gamma}_{_\text{SR}}^{k+1}} \gamma^k G_{1,1}^{1,1}\left(\gamma \Bigg| \begin{array}{l}1-\beta\\0\end{array}\right) G_{0,1}^{1,0}\left(\frac{(\beta_{_\text{SR}} - \delta_{_\text{SR}})}{\bar{\gamma}_{_\text{SR}}} \gamma \Bigg| k-i \right) G^{1,2}_{2,2}\left({\frac{m\Omega_s\gamma}{m_s\Omega_m \bar{\gamma}_{_\text{RU}}}} \middle|\, \begin{gathered}
			1-m_s, 1\\
			m, 0
		\end{gathered}\right)d\gamma \\
		&\overset{(b)}{=} \sum_{k=0}^{\Gamma_{s}-1} \frac{\alpha_{_\text{SR}} c_k}{\Gamma(m)\Gamma(m_s)\Gamma(\beta)\bar{\gamma}_{_\text{SR}}^{k+1}}  G^{1,1:1,0:1,2}_{1,1:0,1:2,2}
		\left(\begin{array}{c|c|cc|} -k& &1-m_s,1&\\  -k-1+\beta&0&m,0&\\\end{array} \frac{\beta_{_\text{SR}} -\delta_{_\text{SR}}}{\bar{\gamma}_{_\text{SR}}},\frac{m\Omega_s}{m_s\Omega_m\bar{\gamma}_{_\text{RU}}}\right). \label{EC-AF-N1} 
		\end{aligned}
	\end{equation}	\rule[-7pt]{18.05cm}{0.05em}  
\end{figure*}

Likewise, by substituting $f_{\gamma_{_\text{RU}}}(\gamma) $ and $F_{\gamma_{_\text{SR}}}(\gamma)$ in Eqs.\eqref{PDF-RU}  and \eqref{CDF-SR}, respectively,  $\mathcal{N}_2$ can be expressed as shown in Eq.~\eqref{EC-AF-N2} at the top of next page. To evaluate the first integral term involving the Gamma function term $\Gamma(k+1)$, a closed-form solution is obtained by applying \cite[Eq.~(3.197.3)]{2007Table}, where $g_k=\frac{\alpha_{_\text{SR}}c_k \left(\frac{m\Omega_s}{m_s\Omega_m \bar{\gamma}_{_\text{RU}}}\right)^{m}}{B(m,m_s)\bar{\gamma}_{_\text{SR}}{(\beta_{_\text{SR}}-\delta_{_\text{SR}})^{k+1}}}$ is defined for simplicity. For the second integral term involving the Meiger-G function, the closed-form expression is obtained with the help of \cite[Eq.~(07.34.03.0271.01)]{walfram}, \cite[Eq.~({07.34.03.0228.01})]{walfram}, and \cite [Eq.~(07.34.21.0081.01)]{walfram}, which is similar to the derivation of $\mathcal{N}_1$. Eventually the final analytical result can be obtained by adding the derived results of $\mathcal{N}$, $\mathcal{N}_1$, and $\mathcal{N}_1$ and taking the logarithm.
\begin{figure*}
	\begin{equation}
		\begin{aligned}
		\mathcal{N}_2 &{=}\int_{0}^{\infty} \sum_{k=0}^{\Gamma_{s}-1}
		\frac{\alpha_{_\text{SR}}c_k \left(\frac{m\Omega_s}{m_s\Omega_m \bar{\gamma}_{_\text{RU}}}\right)^{m}}{B(m,m_s)\bar{\gamma}_{_\text{SR}}{(\beta_{_\text{SR}}-\delta_{_\text{SR}})^{k+1}}} \frac{\gamma^{m-1}}{\left(1+\gamma\right)^\beta (1 + \frac{m\Omega_s}{m_s\Omega_m \bar{\gamma}_{_\text{RU}}} \gamma)^{m+m_s}}  \left[ \Gamma(k+1)-G_{1,2}^{2,0}\left(\frac{(\beta_{_\text{SR}}-\delta_{_\text{SR}})\gamma}{\bar{\gamma}_{_\text{SR}}} \Bigg| \begin{array}{c}1 \\ 0, k+1\end{array}\right)\right] d\gamma\\ 
		&{=} \sum_{k=0}^{\Gamma_{s}-1} 
		g_k \left[\Gamma(k+1)B(m,m_s+\beta){}_2 F_1\left(m+m_s,m;m+m_s+\beta;1-\frac{m\Omega_{s}}{m_s\Omega_m\bar{\gamma}_{_\text{RU}}}\right)\right.\\ &~~~~~~~~\left.-\frac{1}{\Gamma(m+m_s)\Gamma(\beta)} G^{1,1:1,1:2,0}_{1,1:1,1:1,2}
		\left(\begin{array}{c|c|cc|} 1-m&1-m-m_s&1&\\  \beta-m&0&0,k+1& \\ \end{array} \frac{m\Omega_s}{m_s\Omega_m\bar{\gamma}_{_\text{RU}}}, \frac{\beta_{_\text{SR}} -\delta_{_\text{SR}}}{\bar{\gamma}_{_\text{SR}}}\right)\right]. \label{EC-AF-N2} 
		\end{aligned}
	\end{equation}  
	\rule[-7pt]{18.05cm}{0.05em}  
\end{figure*}

\begin{proposition}
The asymptotic expression for the effective capacity of DF relay, denoted by $\widetilde{\mathrm{E}}_{\mathrm{C}}^{\text{DF}}(\theta)$, can be expressed as follows:
\begin{equation}
	\begin{aligned}
	\widetilde{\mathrm{E}}_{\mathrm{C}}^{\text{DF}}(\theta) =& -\frac{1}{\theta} \log \left\{\beta \left[ \sum_{k=0}^{\Gamma_{s}-1} \frac{\alpha_{_\text{SR}}c_k}{(k+1)\bar{\gamma}_{_\text{SR}}^{k+1}} B(k+2, \beta -k-1)\right.\right.\\ &+ \left.\left.\frac{\Gamma(\beta-m)\Gamma(m+m_s)}{\Gamma(\beta+1)\Gamma(m_s)}\left(\frac{m\Omega_s}{m_s\Omega_m \bar{\gamma}_{_\text{RU}}}\right)^{m}\right] \right\}.\label{asy-EC-DF}
	\end{aligned}
\end{equation}\end{proposition}

\begin{IEEEproof} 
Due to the complexity of the derived $\mathrm{E}_{\mathrm{C}}^{\text{DF}}(\theta)$, we opt for a more convenient approach by leveraging its asymptotic representation in high SNR regimes in the following. Plugging Eq.~\eqref{asy-DF-final} into Eq.~\eqref{EC-AF}, we have 
\begin{equation}
	\begin{aligned}
	&\widetilde{\mathrm{E}}_{\mathrm{C}}^{\text{DF}}(\theta) = -\frac{1}{\theta} \log \left\{\beta \int_{0}^{\infty} \frac{1}{(1+\gamma)^{\beta+1}} \times\right.\\ &\left. \left[\sum_{k=0}^{\Gamma_{s}-1}\frac{\alpha_{_\text{SR}}c_k}{(k+1)} \left(\frac{\gamma}{\bar{\gamma}_{_\text{SR}}}\right)^{k+1}+\frac{m^{m-1}\Omega_s^m \gamma^m}{B(m,m_s)(m_s\Omega_m\bar{\gamma}_{_\text{RU}})^m}\right]  d\gamma\right\}.
	\end{aligned}
\end{equation}
For the integral involving $F_{\gamma_{_\text{SR}}}^{\text{asy}}(\gamma)$ and $F_{\gamma_{_\text{RU}}}^{\text{asy}}(\gamma)$, the closed-form expression for $\widetilde{\mathrm{E}}_{\mathrm{C}}^{\text{DF}}(\theta)$ is derived as Eq.~\eqref{asy-EC-DF} with the help of \cite[Eq.~(3.194.3)]{2007Table}, which completes the proof. 
\end{IEEEproof} 
\subsection{Outage Probability and $\epsilon${-}Outage Capacity}\label{op-section}
The outage probability, denoted by $\mathbb{P}$, is defined as the probability that the instantaneous end-to-end SNR falls below a predetermined outage threshold $\gamma_{\text{th}}$. Mathematically, this can be expressed as 
\begin{equation}
\mathbb{P}=\text{Pr}\left[\gamma<\gamma_{\text{th}}\right]=F_{\gamma}\left(\gamma_{\text{th}}\right). 
\end{equation}
Leveraging the expressions of CDF derived in Section \ref{sec3}, i.e., Eqs.~\eqref{ASY-CDF-SR}, \eqref{ASY-CDF-RU}, \eqref{asy-AF-final}, and \eqref{asy-DF-final}, the outage probabilities of the S-R link, R-U link, fixed-AF cascaded link, and DF cascaded link, denoted by $\mathbb{P}_{_\text{SR}}$, $\mathbb{P}_{_\text{RU}}$, $\mathbb{P}_{_\text{AF}}$, and $\mathbb{P}_{_\text{DF}}$, respectively, can be directly obtained as follows:
\begin{equation}
\begin{cases}
~~\mathbb{P}_{_\text{SR}}=F_{\gamma_{_\text{SR}}}\left(\gamma_{\text{th}}\right),~~~~~~\mathbb{P}_{_\text{RU}}=F_{\gamma_{_\text{RU}}}\left(\gamma_{\text{th}}\right),~~~~\\
~~\mathbb{P}_{_\text{AF}}=F_{\gamma_{_\text{AF}}}\left(\gamma_{\text{th}}\right),~~~~~~\mathbb{P}_{_\text{DF}}=F_{\gamma_{_\text{DF}}}\left(\gamma_{\text{th}}\right).\textbf{}  
\end{cases}
\end{equation}

The diversity order, denoted by $G_d$, is an important asymptotic metric that provides deeper insights into understanding and evaluating system performance. It is defined as 
\begin{equation}
\begin{aligned}
G_{d}=-\lim _{\bar{\gamma} \rightarrow \infty} \frac{\log \left(\mathbb{P}^{\text{asy}}\right)}{\log (\bar{\gamma})},
\end{aligned}
\end{equation}
where $\mathbb{P}^{\text{asy}}$ denotes the asymptotic outage probability at high SNR regimes. Likewise, tight asymptotic expressions of outage probability, denoted by $\mathbb{P}_{_\text{SR}}^{\text{asy}}$, $\mathbb{P}_{_\text{RU}}^{\text{asy}}$, $\mathbb{P}_{_\text{AF}}^{\text{asy}}$, and $\mathbb{P}_{_\text{DF}}^{\text{asy}}$, respectively, can be derived by plugging the asymptotic CDFs, i.e., Eqs.~\eqref{CDF-SR},\eqref{CDF-RU},\eqref{CDF-AF}, and \eqref{CDF-DF}, as follows: 
\begin{equation}
\begin{cases}
~~\mathbb{P}_{_\text{SR}}^{\text{asy}}=F_{\gamma_{_\text{SR}}}^{\text{asy}}\left(\gamma_{\text{th}}\right),~~~~~~\mathbb{P}_{_\text{RU}}^{\text{asy}}=F_{\gamma_{_\text{RU}}}^{\text{asy}}\left(\gamma_{\text{th}}\right),~~~~\\
~~\mathbb{P}_{_\text{AF}}^{\text{asy}}=F_{\gamma_{_\text{AF}}}^{\text{asy}}\left(\gamma_{\text{th}}\right),~~~~~~\mathbb{P}_{_\text{DF}}^{\text{asy}}=F_{\gamma_{_\text{DF}}}^{\text{asy}}\left(\gamma_{\text{th}}\right). 
\end{cases}
\end{equation}


For the S-R link, as $\bar{\gamma}_{_\text{SR}} \rightarrow \infty$, the term with $k=0$ dominates in Eq.~\eqref{ASY-CDF-SR}, allowing the omission of the remaining components. Thus, the diversity order of the S-R link can be calculated as $G_d^{_\text{SR}}= 1$. Similarly, The diversity order of the R-U link is $G_d^{_\text{RU}}= m$. 
For the considered relaying system, with $\bar{\gamma}_{_\text{SR}}=\bar{\gamma}_{_\text{RU}}=\bar{\gamma}\rightarrow \infty$, the diversity orders for the fixed-gain AF and DF relays are given by $G_d^{_\text{AF}}=1$ and $G_d^{_\text{DF}}=\min \left\{{1,m}\right\} $, respectively. 

The $\epsilon$-outage
capacity, denoted by $C_\epsilon$, represents the largest transmission rate for which the outage probability $\mathbb{P}$ is less than $\epsilon$. Mathematically, it is defined as
\begin{equation}
C_\epsilon=\sup\left\{R~|~\text{Pr}\left[\gamma<\gamma_{\text{th}}\right]\le \epsilon\right\}, 
\end{equation}
where $\sup\left\{\cdot\right\}$ denotes the supremum operator.

Based on the closed-form outage probability expressions derived above, $C_\epsilon$ can be computed by solving the following equation for a given $\epsilon$ as 
\begin{equation}
\mathbb{P} = F_{\gamma}(2^{C_\epsilon} - 1) = \epsilon.
\end{equation}
For various links, the solutions to corresponding transcendental equations are achievable through numerical solvers in MATLAB, such as the \texttt{fzero} and \texttt{fsolve} functions, which provide rapid convergence with precision-controlled tolerance settings.
\begin{corollary}
	An important limiting case of the $\epsilon$-outage capacity is the \emph{zero-outage capacity}, corresponding to $\epsilon \to 0^+$. In this regime, the outage-free transmission rate becomes equivalent to the \emph{delay-limited capacity}, which also emerges as the asymptotic limit of the effective capacity under increasingly stringent QoS constraints, i.e., as $\theta \to \infty$\cite{outagecapacity}. Formally,
	\begin{equation}
	\lim_{\epsilon \to 0^+} C_\epsilon = \lim_{\theta \to \infty} EC(\theta).
	\end{equation}
	This convergence implies that both the zero-outage capacity and the delay-limited capacity characterize the maximum sustainable rate under ultra-reliable and extremely strict delay constraints.
	For i.i.d. continuous fading channels, it is well known that the zero-outage capacity converges to 0.
\end{corollary}

\section{Numerical Results and Discussions}\label{sec5}
In this section, we first implement the field measurement of the proposed Fisher-Snedecor $\mathcal{F}$ channel as a supplement to \cite{tcomF}. The applicability of the proposed Fisher-Snedecor $\mathcal{F}$ distribution for modeling complex channel characteristics in post-disaster scenarios is demonstrated through comparative analysis with empirical data against conventional fading models.
Subsequently, we validate the derived analytical expressions through numerical evaluations, supported by Monte Carlo simulations to validate the effectiveness of the analysis.
Unless otherwise specified, the main parameters are given as follows: The fading parameter of S-R link is set to $\Gamma_{s}=2$ and the average power parameters are set to $\Omega_{sr}=2 b_s=0.5$ to  normalize the mean power of S-R link as $\mathbb{E}[\left|h_{_\text{SR}}\right|^{2}]=\Omega_{sr}+2 b_s=1$.
The average power parameters of R-U link are $\Omega_s^2={m_s}/{(m_s-1)}$ and $\Omega_m^2={(m_s-1)}/{m_s}$, which are set to normalize the mean power of R-U link as $\mathbb{E}[\left|h_{_\text{RU}}\right|^{2}]=\Omega_m\Omega_s=1$. 
The satellite and UAV relay maintain respective altitudes of 765 km and 150 m, while path loss parameters are set as $\alpha_p=2$, $\eta_{\text{LoS}}=0.1$ dB, $\eta_{\text{NLoS}}=20$ dB. The communication system operates with SNR threshold  $\gamma_\text{th}=$ 10 dB and noise power $\sigma^2_{_\text{SR}}=\sigma^2_{_\text{RU}}$= -94 dBm, where both S-R and R-U links maintain identical average SNRs  ${\bar{\gamma}_{_\text{SR}}}={\bar{\gamma}_{_\text{RU}}}={\bar{\gamma}}$. \textcolor{black}{The fixed-AF relaying gain $G = 1$.} The value of $B$ and $T$ are set to 20 MHz and 2 ms, respectively\cite{Bandwidth,T}.
\begin{figure}
	\centering
	\includegraphics[width=.98\columnwidth]{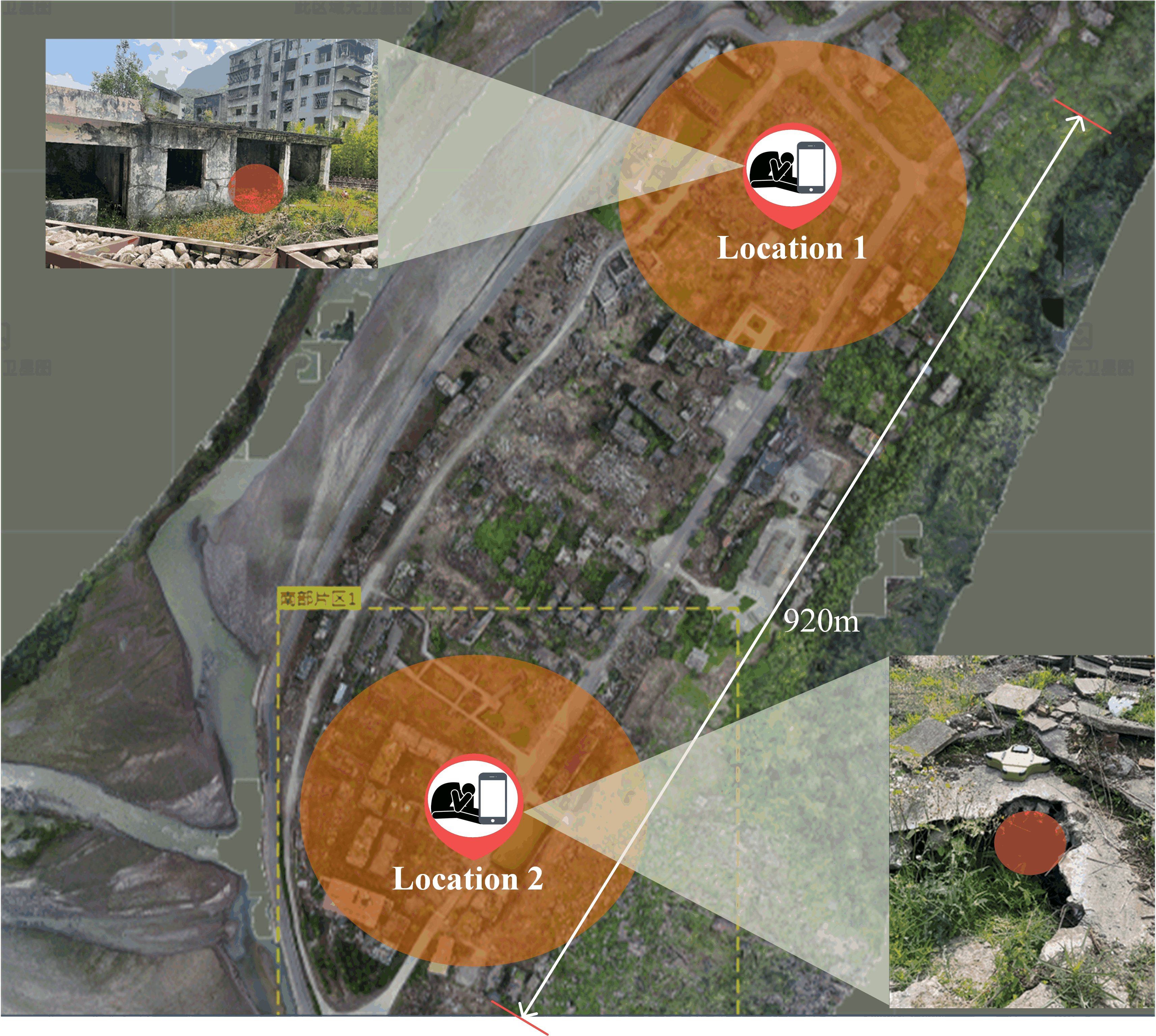}
	\caption{Field measurement site where the UE was placed in 1) partially collapsed room corner and 2) collapsed underground space.}\label{measure}\vspace{-2pt}
\end{figure}

\subsection{Measurement Setup and Fading Model Evaluation}

To characterize post-disaster communication channels, field measurements were conducted at Beichuan Earthquake Relics Site, China, a location preserving typical disaster-induced geological features including collapsed mountains, debris flows, dammed lakes, and various collapsed infrastructure. The experimental framework considered ground-to-ground (G2G) communications which simulated rescue-team-to-victim interactions. As shown in Fig.~\ref{measure}, two measurement locations were selected to represent different levels of obstruction:
(1) A moderately obstructed location, where the UE was placed in a partially collapsed room corner, with structural remnants and debris causing intermittent or reflected signal paths;
(2) A severely obstructed location, where the UE was positioned in a collapsed underground space, resulting in significant signal attenuation due to surrounding debris and construction materials.

For all of the measurements performed in this study,  transceiver equipment used is USRP-2954, operating at 1450MHz.
The antennas used by the transmitter and receiver were Taoglas FXUB63 flexible PCB antennas. 
To emulate the UE (mobile phone), the packaging step of receiving antenna was adopted as described in \cite{channelmeasurement}.
First, the UE is vertically placed at two selected locations, respectively. A three-dimensional coordinate system is then established with the stationary UE positioned at ground zero, serving as the spatial reference origin. Second, the transmitter systematically traverses a 100m radius circular area to collect 650 samples of the received signal power across predetermined grid points. 
After the data acquisition, the signal processing pipeline involves three sequential stages. The log-distance model $PL(d) = 10\alpha \log_{10}(d/d_0)$ is applied for the path loss compensation process, where the path loss exponent $\alpha = 2.7$ is empirically determined from the measurement data. Subsequently, the normalized signal envelope is analyzed through empirical PDF and CDF estimations. Ultimately, statistical modeling is performed by fitting Rayleigh, Nakagami-$m$, and Fisher-Snedecor $\mathcal{F}$ distributions via nonlinear least-squares optimization, with model selection criteria based on mean squared error (MSE) quantification.

\begin{figure}
	\centering
	\includegraphics[width=1.05\columnwidth]{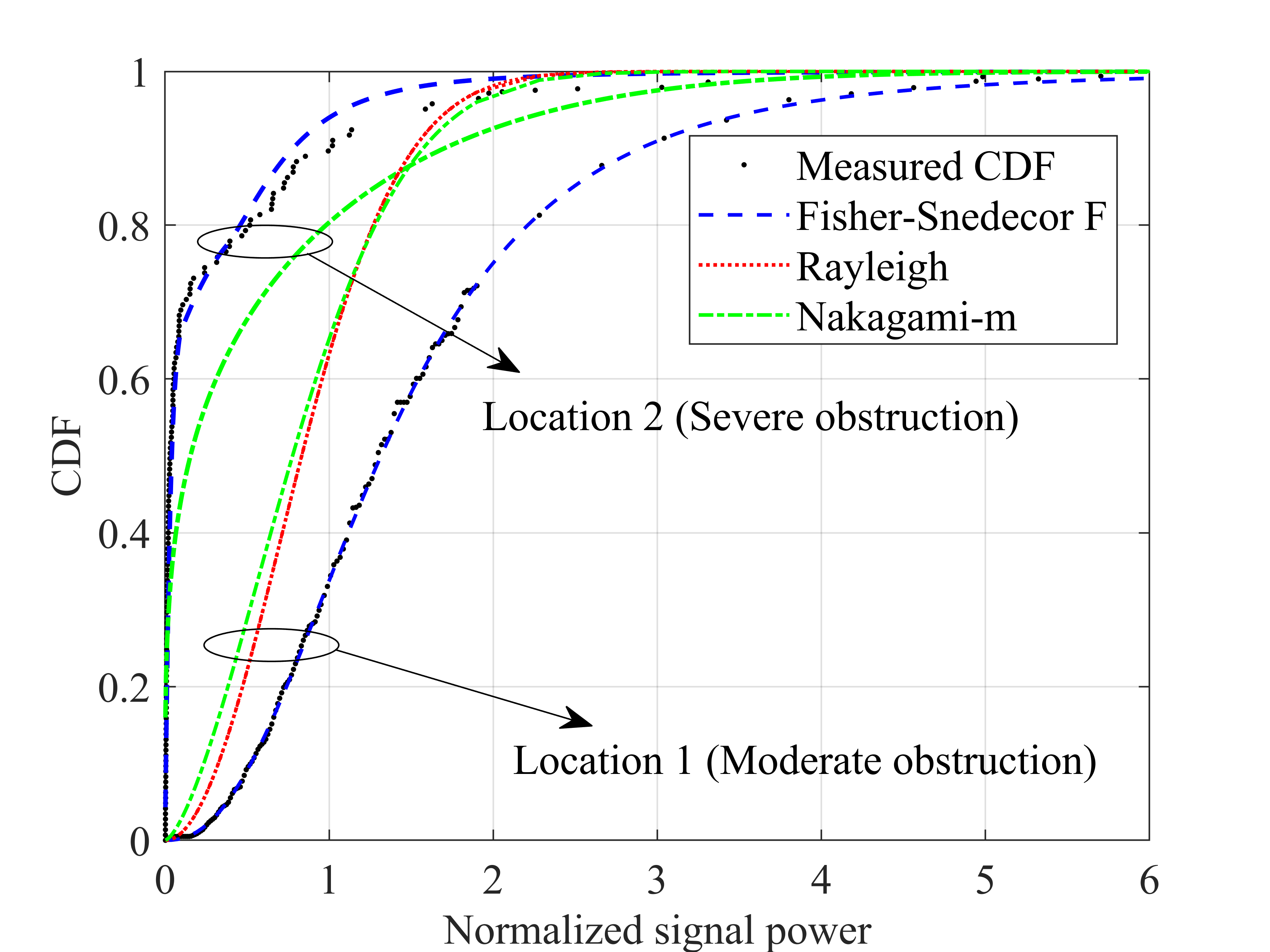}
	\caption{CDFs of the measured data and the fitted curves using three models: Rayleigh, Nakagami-$m$, and Fisher-Snedecor $\mathcal{F}$ distribution.}
	\label{CDFcomparison}
\end{figure}

All parameter estimates given in Table \ref{est} demonstrate the superior fitting accuracy of the Fisher-Snedecor $\mathcal{F}$ model across both measurement scenarios. Fig.~\ref{CDFcomparison} presents the CDF comparison of the Rayleigh, Nakagami-$m$, and Fisher-Snedecor $\mathcal{F}$ fading models fitted to empirical data. 
Specially, at location 1, both the Rayleigh and Nakagami-$m$ models exhibit significant deviations from the empirical curve, whereas the Fisher-Snedecor $\mathcal{F}$ distribution provides a remarkably close fit across the entire signal power range.
At location 2, the Fisher-Snedecor $\mathcal{F}$ model offers the best fit throughout the full CDF range. Although the Rayleigh and Nakagami-$m$ models approximate the tail behavior (high signal power region) relatively well, clear mismatches exist in the low and medium power regions.  
From the perspective of MSE, the $\mathcal{F}$ distribution outperforms the Rayleigh and Nakagami-$m$ models by factors of 60 and 562 at Location 1, respectively. More strikingly at location 2, the $\mathcal{F}$ model achieves a MSE that is two orders of magnitude lower than that of the competing models. The above analysis indicates that in harsh and obstructed post-disaster environments, the $\mathcal{F}$ distribution provides superior fitting performance to the channel characterizations with better flexibility and compatibility. 


\begin{table*}[htbp]
	\centering
	\caption{PARAMETER ESTIMATES FOR THE STATISTICAL CHANNEL MODELS FOR ALL OF THE CONSIDERED EWC MEASUREMENTS ALONG WITH THE MSE RESULTS}
	\label{tab:ewc_parameters}
	\begin{tabular*}{\textwidth}{@{\extracolsep{\fill}}|c|ccccc|ccc|cc|}
		\hline
		\multirow{2}{*}{\textbf{EWC scenarios}} 
		& \multicolumn{5}{c|}{\textbf{$\mathcal{F}$ Distribution} ($m$, $m_s$, $\Omega_m$, $\Omega_s$)} 
		& \multicolumn{3}{c|}{\textbf{Nakagami} ($m$, $\Omega$)} 
		& \multicolumn{2}{c|}{\textbf{Rayleigh} ($s$)} \\
		
		& $m$ & $m_s$ & $\Omega_m$ & $\Omega_s$ & MSE &  $m$ & $\Omega$ & MSE & $s$ & MSE \\
		\hline

		Location 1 (Moderate obstruction)& 1.0721  & 2.0211 & 0.7107 & 1.4069 & 3.30e-5 & 0.7453 & 1.0000 & 1.277e-3 & 0.7071 & 4.181e-3  \\
		Location 2 (Severe obstruction)& 0.1721 & 1.1211 &0.3289& 3.0425 & 3.16e-4 & 0.1309 & 1.0000 & 1.887e-2 & 0.7071 & 1.775e-1 \\
		\hline
	\end{tabular*}\label{est}
\end{table*}

\subsection{Performance analysis}
\begin{figure}
	\centering
	\includegraphics[width=1.06\columnwidth]{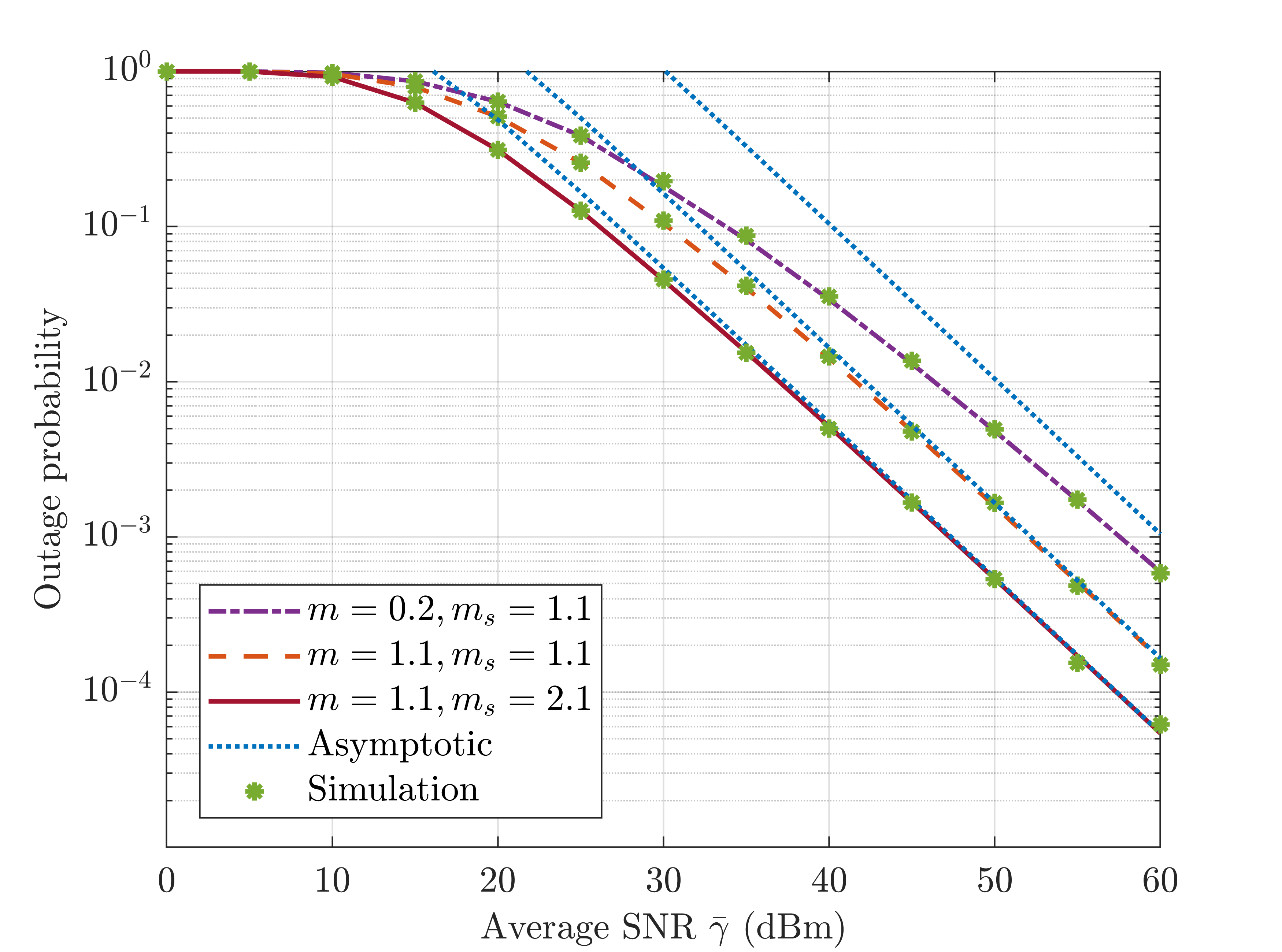}
	\caption{Outage probability of fixed-gain AF relay corresponding to $m$ and $m_s$ versus $\bar{\gamma}$.}
	\label{AF_OP}
\end{figure}
Figure \ref{AF_OP} depicts the outage probability of the fixed-gain AF relaying scheme versus the average SNR $\bar{\gamma}$, under varying composite fading parameters $m$ and $m_s$. As it can be observed, the outage probability deteriorates with more severe multipath fading and shadowing (i.e., lower $m$ and $m_s$), indicating the pronounced impact of the R-U-link channel condition over the overall AF relay system performance. Analytical results align well with Monte Carlo simulations for various fading parameters, thus verifying the effectiveness of the closed-form expressions. 
Moreover, the asymptotic results converge to the exact analysis as  $\bar{\gamma}$ increases,
especially for case of small multipath fading and shadowing parameters. Notably, all asymptotic curves exhibit identical slopes in the high-SNR regime, which confirms that the diversity order of the fixed-gain AF relay system remains invariant with respect to different fading parameters. This consistency reflects the diversity gain of the AF relay system, independent of the composite channel severity.

\begin{figure}
	\centering
	\includegraphics[width=1.06\columnwidth]{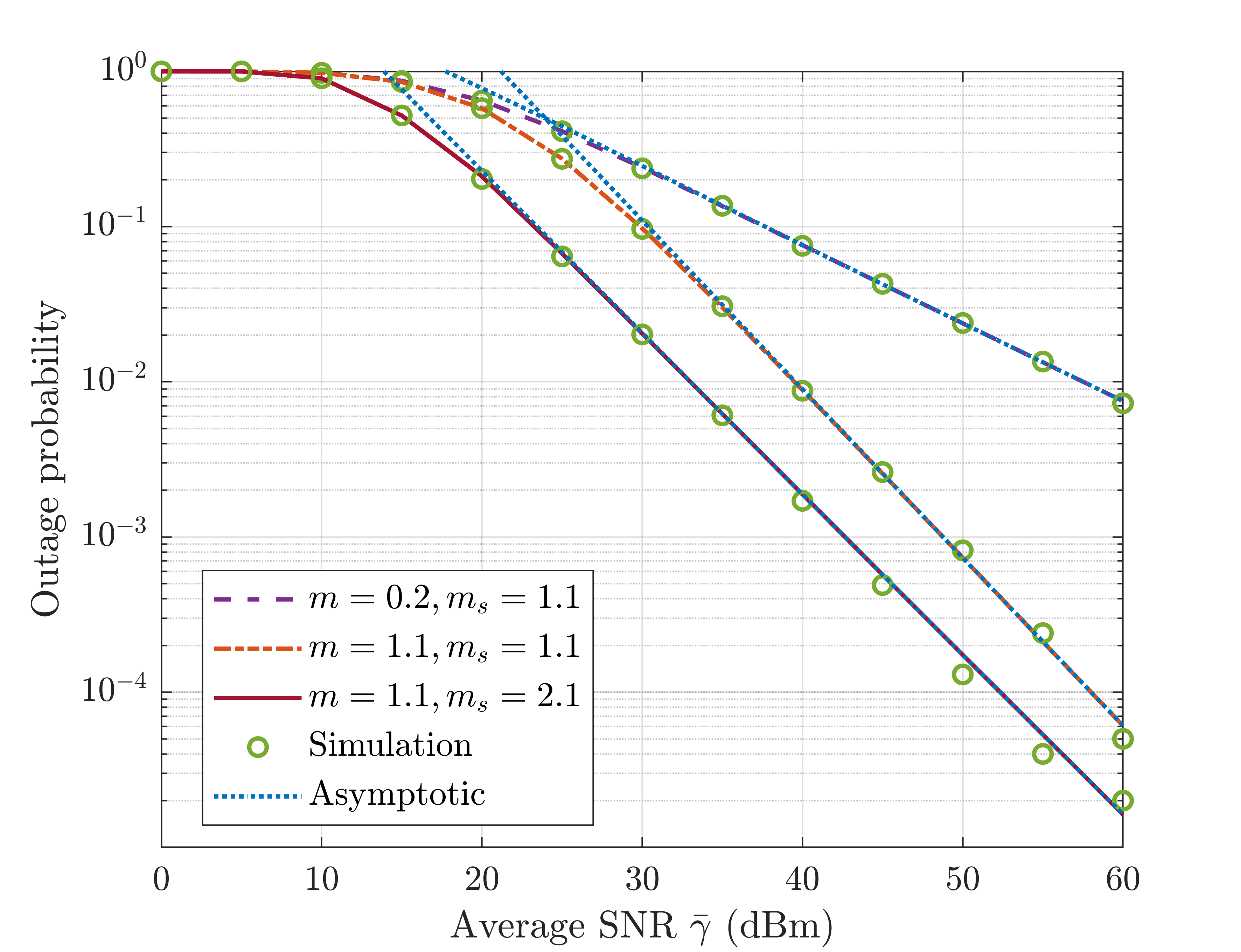}
	\caption{Outage probability of DF relay corresponding to $m$ and $m_s$ versus $\bar{\gamma}$.}
	\label{DF_OP_mmschange}
\end{figure}

\begin{figure}
	\centering
	\includegraphics[width=1.06\columnwidth]{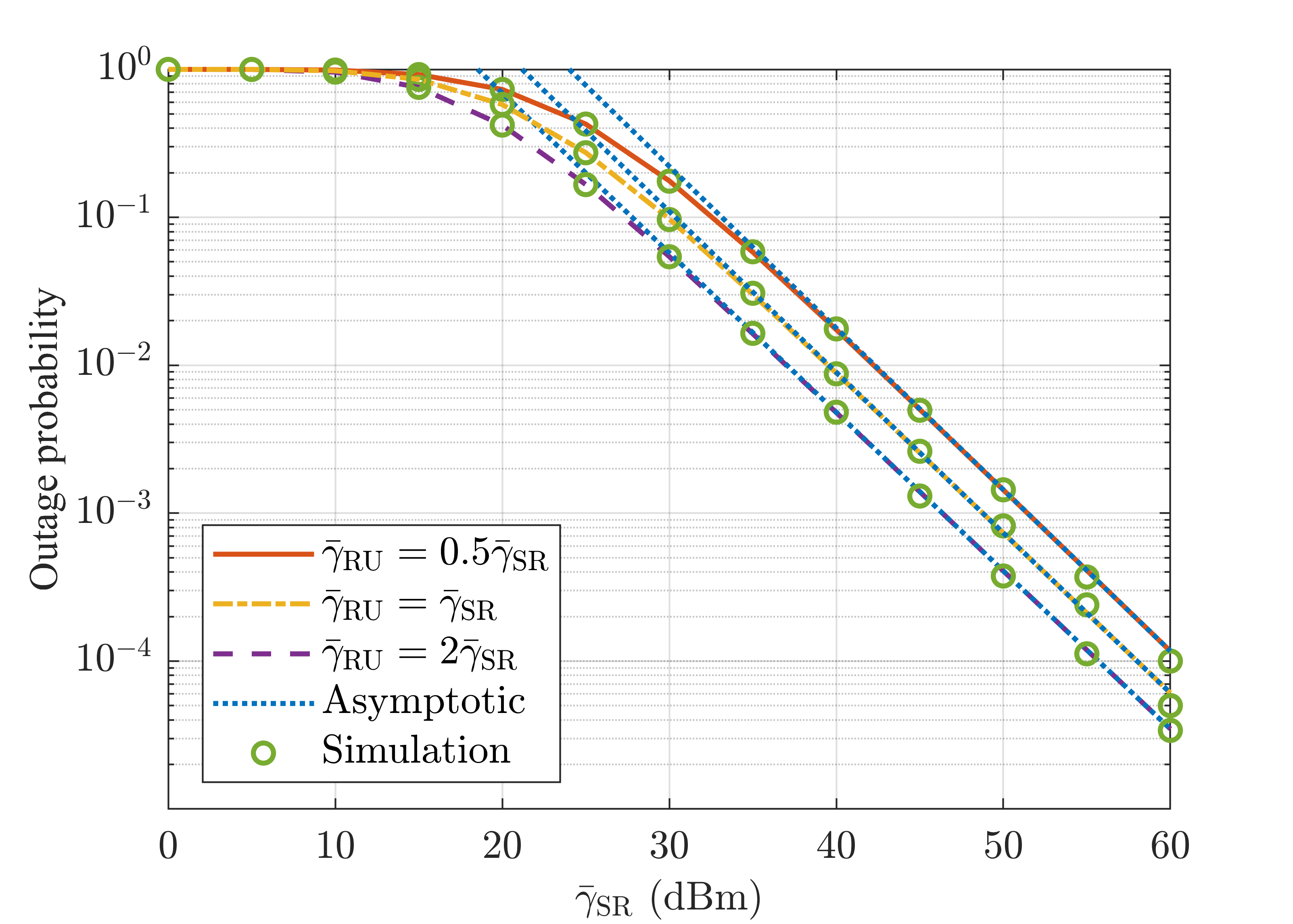}
	\caption{Outage probability of DF relay versus $\bar{\gamma}_\text{SR}$ with $m=1.1, m_s = 1.1$ and $\bar{\gamma}_\text{RU}=\lambda\bar{\gamma}_\text{SR}$, for various values of $\lambda$.}
	\label{DF_OP_snrchange}
\end{figure}

Figure \ref{DF_OP_mmschange} plots the outage probability of DF relaying scheme versus the average SNR $\bar{\gamma}$ with different composite fading parameters $m$ and $m_s$. As expected, the outage probability significantly increases in the presence of harsher multipath fading and shadowing. 
The asymptotic outage probabilities provide an accurate approximation in the medium- and high-SNR regimes, particularly for cases with lighter multipath fading and shadowing. Notably, the asymptotic OP with $m=0.2, m_s=1.1$ exhibits a distinct gradient compared to the other cases, confirming that the diversity gain of the DF relaying scheme is directly influenced by severity of multipath fading. 
Fig.~\ref{DF_OP_snrchange} plots the outage probability of DF relay versus $\bar{\gamma}_\text{SR}$ when $\bar{\gamma}_\text{RU}= \lambda\bar{\gamma}_\text{SR}$ is assumed for various values of $\lambda$ with fading parameters being  $m=1.1, m_s = 1.1$. As depicted, the outage probability consistently decreases with increasing $\lambda$, 
confirming that the quality of the R-U link has a significant impact on the end-to-end reliability of the DF relaying system.
It's noteworthy that since the outage probability of DF relay strongly depends on the R-U link, one can infer that the S-R link exhibits relatively better channel conditions than the R-U link in the considered setting.

\begin{figure}
	\centering
	\includegraphics[width=1.05\columnwidth]{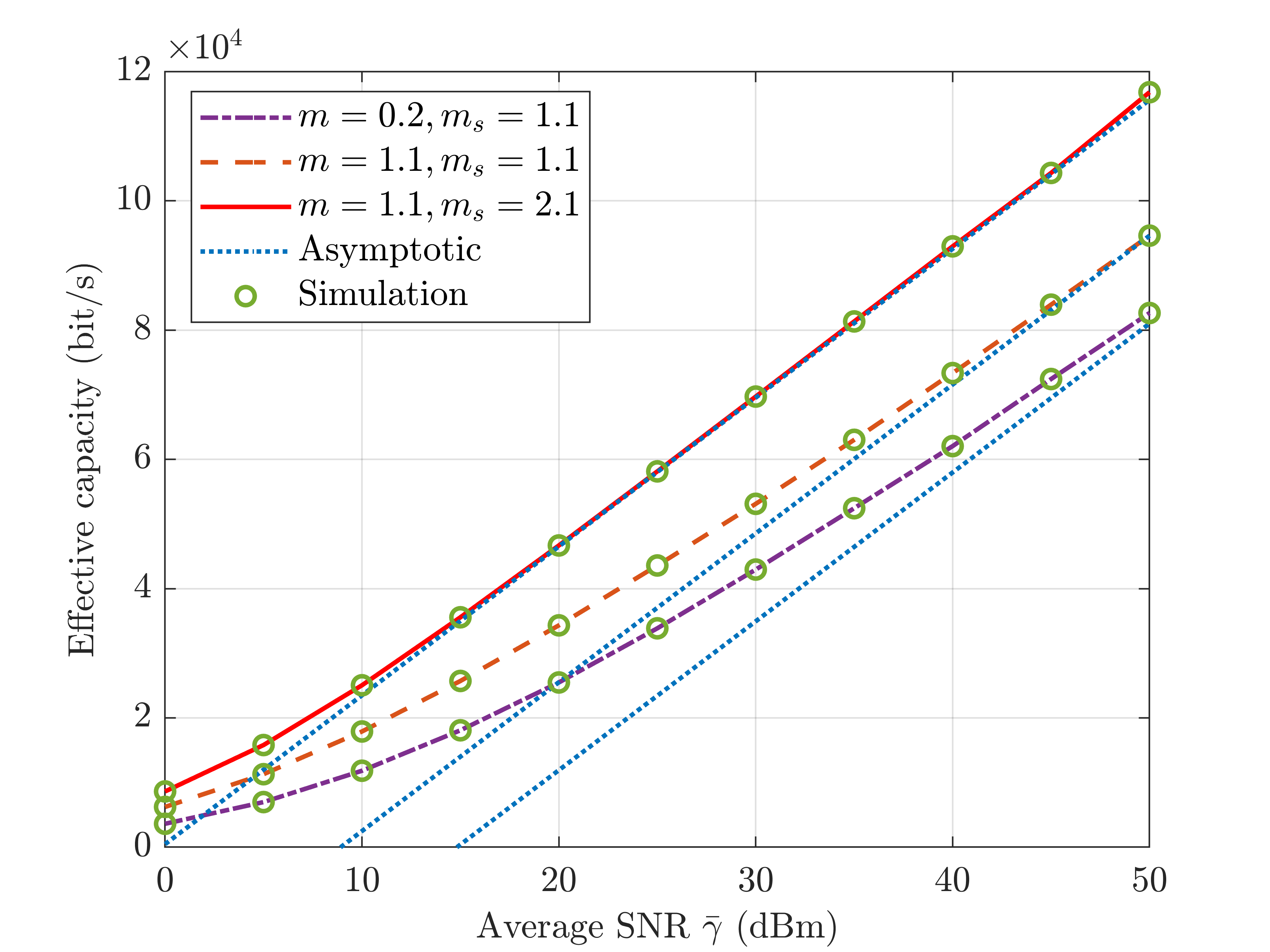}
	\caption{Effective capacity of fixed-gain AF relay corresponding to $m$ and $m_s$ versus $\bar{\gamma}$ with $\theta=1e^{-2}$.}
	\label{AF_EC_mmschange}
\end{figure}

\begin{figure}
	\centering
	\includegraphics[width=1.05\columnwidth]{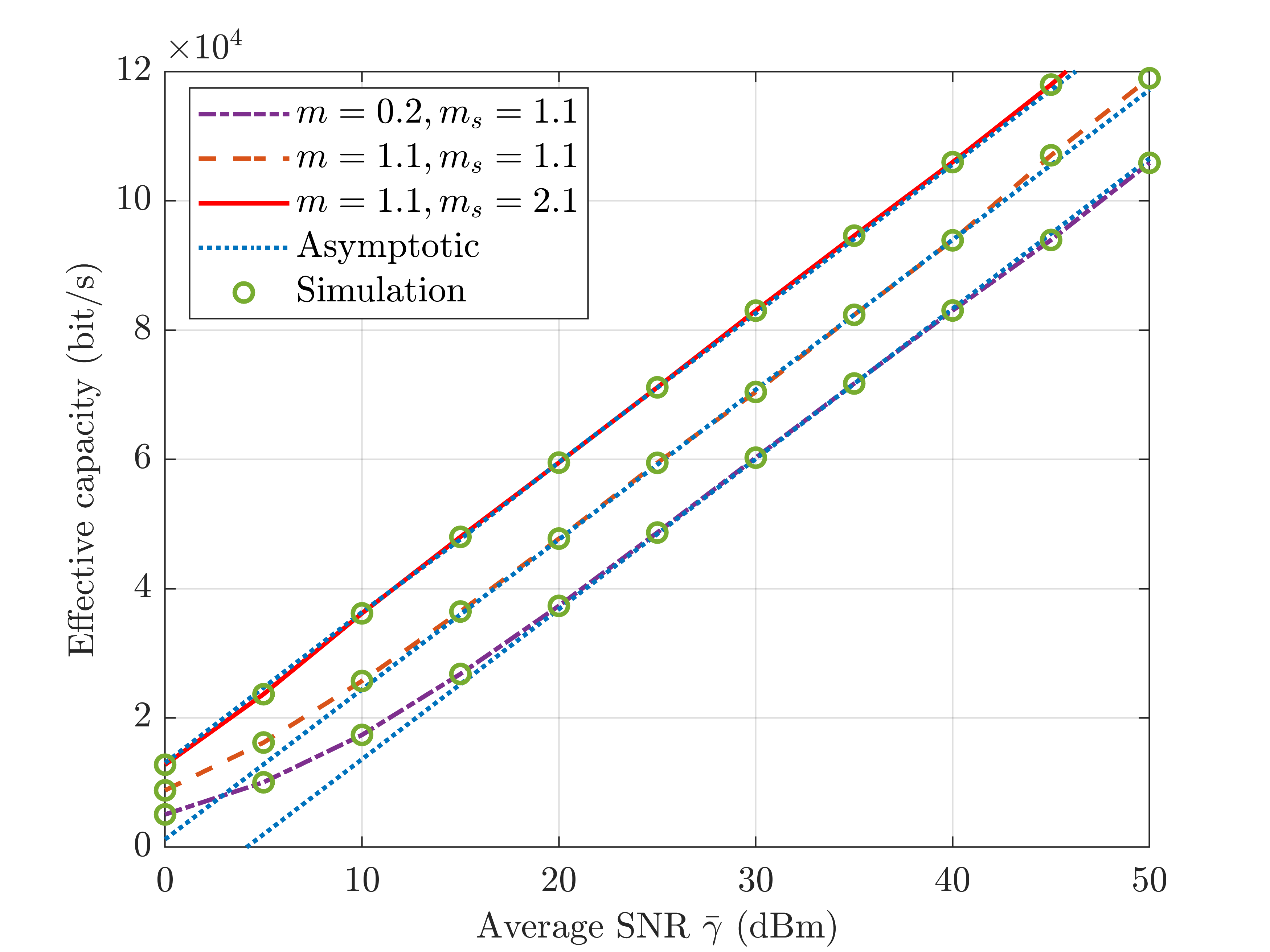}
	\caption{Effective capacity of DF relay corresponding to $m$ and $m_s$ versus $\bar{\gamma}$ with $\theta=1e^{-2}$.}\vspace{-8pt}
	\label{DF_EC_mmschange}
\end{figure}

\begin{figure}
	\centering
	\includegraphics[width=1.1\columnwidth]{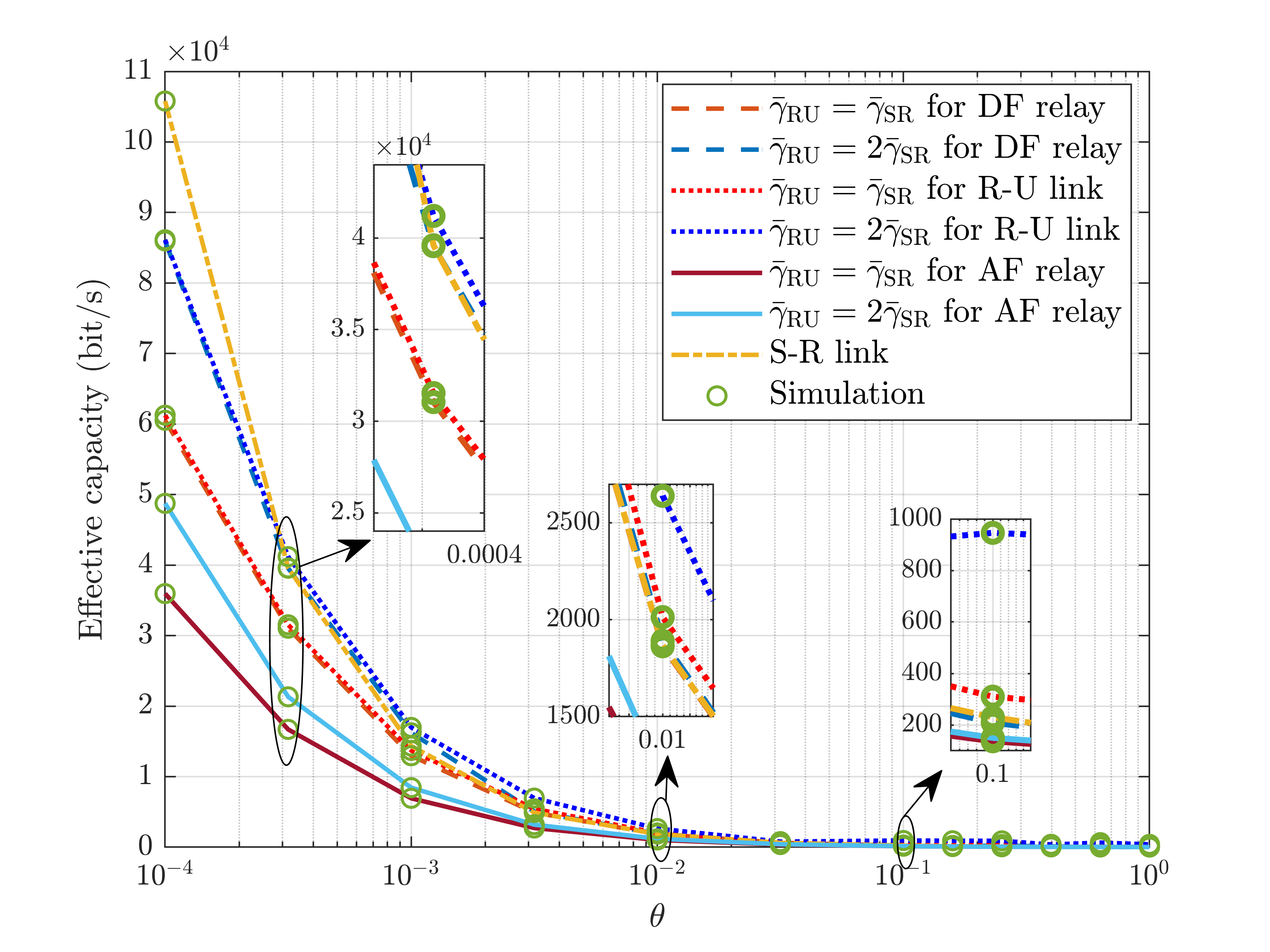}\vspace{-5pt}
	\caption{Effective capacity of various schemes versus $\theta$ with $m=1.1$, $m_s=2.1$, $\bar{\gamma}_{\text{SR}}=30$ dBm, and $\bar{\gamma}_\text{RU}=\lambda\bar{\gamma}_\text{SR}$ for various values of $\lambda$.}\vspace{-2pt}
	\label{DFAFSRRUEC}
\end{figure}
Figures~\ref{AF_EC_mmschange} and \ref{DF_EC_mmschange} present the effective capacity of the fixed-gain AF and DF relaying schemes, respectively, against $\bar{\gamma}$ with different $m$ and $m_s$. As observed, the effective capacity improves with lighter multipath fading and shadowing (i.e., higher $m$ and $m_s$) for both schemes, highlighting the significant influence of the R-U-link's channel condition on the overall system performance. 
Furthermore, one can observe that the DF relay consistently achieves a higher effective capacity than the AF relay, owing to its ability to decode and regenerate the signal at the relay node.
Additionally, the close alignment between the asymptotic and exact results becomes increasingly pronounced in the high-SNR regime, particularly in lighter fading scenarios ($m=1.1, m_s=2.1$) for both systems, confirming the accuracy and practical applicability of the proposed asymptotic expressions for performance evaluation.

Figure~\ref{DFAFSRRUEC} illustrates the effective capacity of various schemes, including the S-R link, R-U link, fixed-gain AF relay, and DF relay, as a function of the QoS exponent $\theta$. The assumption  $\bar{\gamma}_{\text{RU}}= \lambda\bar{\gamma}_{\text{SR}}$ holds for different values of $\lambda$, with the fading parameters being  fixed at $m=1.1$, $m_s=2.1$, and the average SNR of the S-R link being set to $\bar{\gamma}_{\text{SR}}=30$ dBm, respectively. Therefore, the S-R link yields a single curve. It can be observed that when $\theta$ is very small, the effective capacity of the DF relay closely follows that of the R-U link for both considered values of $\lambda$, which is attributed to the dominance of the weaker R-U link due to $\bar{\gamma}_{\text{SR}} > \bar{\gamma}_{\text{RU}}$.
As $\theta$ and $\lambda$ increase, the effective capacity of the DF relay gradually approaches that of the S-R link. This observation highlights that under stricter delay-QoS requirements and higher $\bar{\gamma}_\text{RU}$, the effective capacity of the S-R link decays faster than that of the R-U link, making it the performance floor in DF relaying scheme. 
Notably, at approximately $\theta=10^{-2}$, the effective capacity of the DF relay converges to that of the S-R link for all values of $\lambda$, indicating that beyond this point, the S-R link becomes the limiting factor for the DF relay's performance. Furthermore, the DF relay consistently outperforms the fixed-gain AF relay as expected.
Overall, the effective capacity of the relaying system is jointly influenced by the channel conditions of individual links and the delay-QoS constraint.

\begin{figure}
	\centering
	\includegraphics[width=1.1\columnwidth]{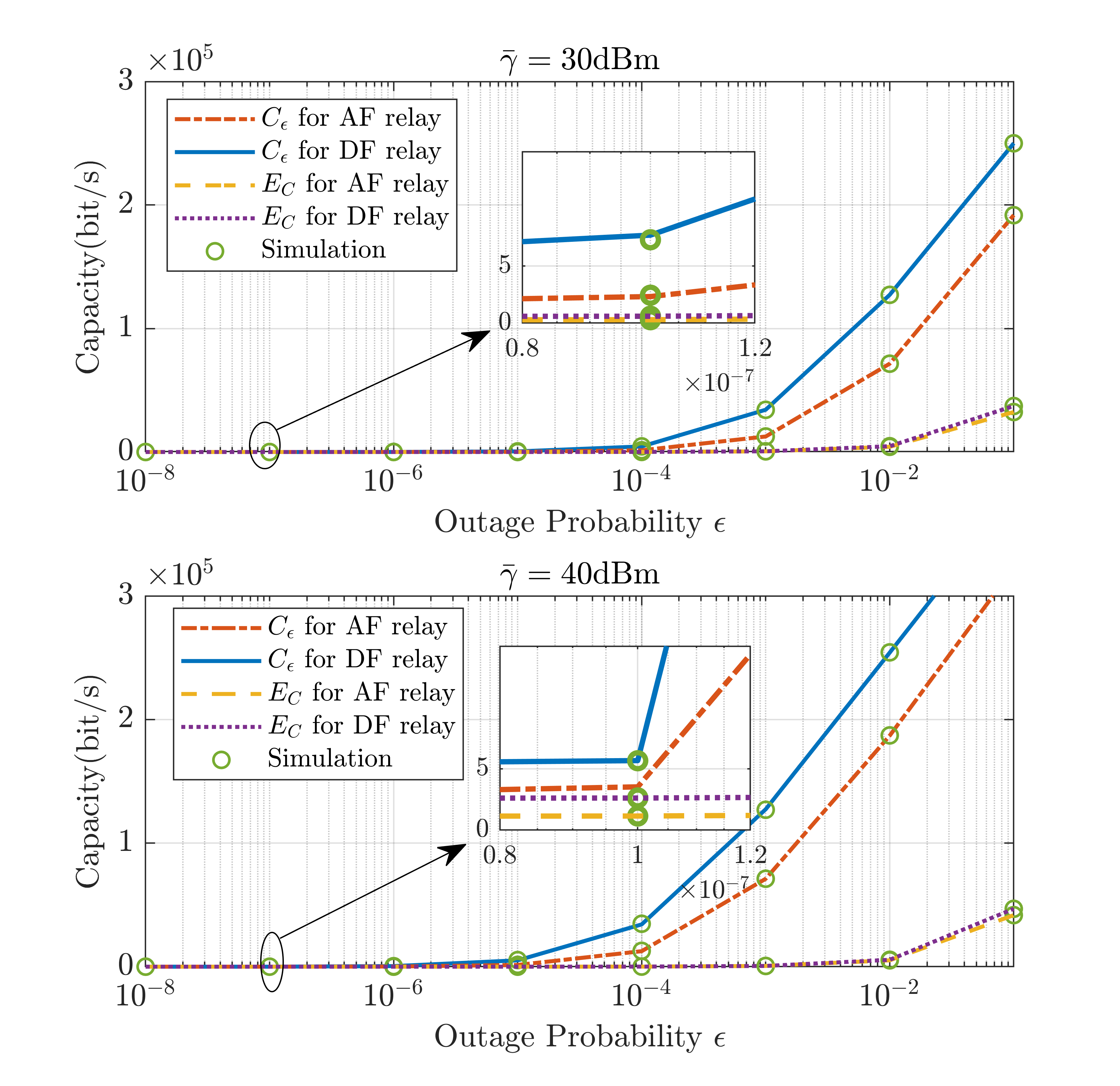}\vspace{-5pt}
	\caption{$\epsilon$-Outage capacity versus $\epsilon$ and effective capacity versus $1/\theta$ of various relaying schemes with $\bar{\gamma}=30$ and $40$ dBm.}\vspace{-2pt}
	\label{Cout_EC_mmschange}
\end{figure}

Figure~\ref{Cout_EC_mmschange} depicts the $\epsilon$-outage capacity as a function of the outage probability $\epsilon$, in comparison with the effective capacity plotted against the reciprocal of the QoS exponent $\theta$ for fixed-gain AF and DF relaying schemes. Here, $\theta = 1/\epsilon$ is set to ensure consistent delay stringency scaling. The average SNR is set to $\bar{\gamma} = {30, 40}$ dBm, respectively. 
It can be clearly observed that as $\epsilon \to 0^+$ and equivalently $\theta \to \infty$, both the $\epsilon$-outage capacity and effective capacity approach the same limiting value $0$. This convergence is evident from the fact that when $\epsilon = 10^{-7}$, the capacities of all schemes are already close to zero. This confirms the theoretical equivalence between the zero-outage capacity and delay-limited capacity under stringent delay or reliability requirements, as established in Corollary~1. Notably, the convergence trend becomes more pronounced at higher SNR (e.g., $\bar{\gamma} = 40$ dB). In this case, the gap between the $\epsilon$-outage capacity and effective capacity diminishes more rapidly, and both capacities stabilize near their common asymptotic limit under favorable channel conditions. 
Similarly, the DF relaying scheme consistently achieves higher capacities than AF scheme for both outage capacity and effective capacity as expected.  
\section{Conclusions}\label{sec6}
To tackle the problem of the mismatch between the theoretical analysis and real performance for SAGIN-enabled EWC systems, in this paper we proposed performance modeling schemes for the SAGIN system by utilizing the composed Fisher-Snedecor $\mathcal{F}$ distribution in the presence of QoS constraints. We developed the exact distribution of the end-to-end signal-to-noise (SNR) statistics for both the space-air and air-ground links. These distributions are then used to analyze the performance of the cascaded link under both fixed-gain AF and DF relaying protocols. Also, we provided the asymptotic expressions of the newly derived results in the high-SNR regime, exhibiting favorable alignment. Furthermore, we investigated the insightful closed-form and asymptotic expressions of effective capacity with QoS provisioning, outage probability, and $\epsilon$-outage capacity. 
Field measurements of the proposed Fisher-Snedecor $\mathcal{F}$ channel were implemented to validate its superior fitting performance. Finally,  Monte Carlo simulations were provided to validate the accuracy and practical relevance of our analytical framework.

\bibliographystyle{IEEEtran}
\bibliography{Ref}

\begin{thebibliography}{10}
\providecommand{\url}[1]{#1}
\csname url@samestyle\endcsname
\providecommand{\newblock}{\relax}
\providecommand{\bibinfo}[2]{#2}
\providecommand{\BIBentrySTDinterwordspacing}{\spaceskip=0pt\relax}
\providecommand{\BIBentryALTinterwordstretchfactor}{4}
\providecommand{\BIBentryALTinterwordspacing}{\spaceskip=\fontdimen2\font plus
\BIBentryALTinterwordstretchfactor\fontdimen3\font minus
  \fontdimen4\font\relax}
\providecommand{\BIBforeignlanguage}[2]{{%
\expandafter\ifx\csname l@#1\endcsname\relax
\typeout{** WARNING: IEEEtran.bst: No hyphenation pattern has been}%
\typeout{** loaded for the language `#1'. Using the pattern for}%
\typeout{** the default language instead.}%
\else
\language=\csname l@#1\endcsname
\fi
#2}}
\providecommand{\BIBdecl}{\relax}
\BIBdecl

\bibitem{EWCUAV}
D.~{G. C.}, A.~Ladas, Y.~A. Sambo, H.~Pervaiz, C.~Politis, and M.~A. Imran,
  ``An overview of post-disaster emergency communication systems in the future
  networks,'' \emph{IEEE Wireless Communications}, vol.~26, no.~6, pp.
  132--139, 2019.

\bibitem{Harsh}
Z.~Yao, W.~Cheng, W.~Zhang, T.~Zhang, and H.~Zhang, ``The rise of {UAV} fleet
  technologies for emergency wireless communications in harsh environments,''
  \emph{IEEE Network}, vol.~36, no.~4, pp. 28--37, 2022.

\bibitem{ScienceChina}
T.~Zhang, C.~Chen, Y.~Xu, J.~Loo, and W.~Xu, ``Joint task scheduling and
  multi-uav deployment for aerial computing in emergency communication
  networks,'' \emph{Science China Information Sciences}, vol.~66, no.~9, p.
  192303, 2023.

\bibitem{Yao}
Z.~Yao, W.~Cheng, W.~Zhang, and H.~Zhang, ``Resource allocation for
  {5G-UAV-based} emergency wireless communications,'' \emph{IEEE Journal on
  Selected Areas in Communications}, vol.~39, no.~11, pp. 3395--3410, 2021.

\bibitem{UAVBS1}
M.~M. Azari, F.~Rosas, K.-C. Chen, and S.~Pollin, ``Ultra reliable uav
  communication using altitude and cooperation diversity,'' \emph{IEEE
  Transactions on Communications}, vol.~66, no.~1, pp. 330--344, 2018.

\bibitem{UAVrelay1}
L.~Yang, J.~Chen, M.~O. Hasna, and H.-C. Yang, ``Outage performance of
  {UAV}-assisted relaying systems with {RF} energy harvesting,'' \emph{IEEE
  Communications Letters}, vol.~22, no.~12, pp. 2471--2474, 2018.

\bibitem{UAVrelay2}
B.~Ji, Y.~Li, D.~Cao, C.~Li, S.~Mumtaz, and D.~Wang, ``Secrecy performance
  analysis of uav assisted relay transmission for cognitive network with energy
  harvesting,'' \emph{IEEE Transactions on Vehicular Technology}, vol.~69,
  no.~7, pp. 7404--7415, 2020.

\bibitem{UAVshorts}
N.~Cheng, W.~Xu, W.~Shi, Y.~Zhou, N.~Lu, H.~Zhou, and X.~Shen, ``Air-ground
  integrated mobile edge networks: Architecture, challenges, and
  opportunities,'' \emph{IEEE Communications Magazine}, vol.~56, no.~8, pp.
  26--32, 2018.

\bibitem{SAGIN_intro1}
M.~Wu, K.~Guo, X.~Li, M.~Asif, C.~Han, and K.~M. Rabie, ``{URLLC} for
  {DRL-RIS-AIDED SAGIN}s: New mentalities, trends and preliminary solutions,''
  \emph{IEEE Communications Standards Magazine}, vol.~9, no.~1, pp. 6--12,
  2025.

\bibitem{SAGIN_intro2}
H.~Cui, J.~Zhang, Y.~Geng, Z.~Xiao, T.~Sun, N.~Zhang, J.~Liu, Q.~Wu, and
  X.~Cao, ``Space-air-ground integrated network ({SAGIN}) for 6{G: }
  requirements, architecture and challenges,'' \emph{China Communications},
  vol.~19, no.~2, pp. 90--108, 2022.

\bibitem{SAGIN_intro3}
X.~Yuan, F.~Tang, M.~Zhao, and N.~Kato, ``Joint rate and coverage optimization
  for the thz/rf multi-band communications of space-air-ground integrated
  network in 6g,'' \emph{IEEE Transactions on Wireless Communications},
  vol.~23, no.~6, pp. 6669--6682, 2024.

\bibitem{SAGIN-NAKAM}
M.~Wu, K.~Guo, Z.~Lin, X.~Li, K.~An, and Y.~Huang, ``Joint optimization design
  of {RIS}-assisted hybrid {FSO} {SAGIN}s using deep reinforcement learning,''
  \emph{IEEE Transactions on Vehicular Technology}, vol.~73, no.~3, pp.
  3025--3040, 2024.

\bibitem{saginshannon}
S.~Wang, L.~Yang, X.~Li, K.~Guo, H.~Liu, H.~Song, and R.~H. Jhaveri,
  ``Performance analysis of satellite-vehicle networks with a non-terrestrial
  vehicle,'' \emph{IEEE Transactions on Intelligent Vehicles}, vol.~9, no.~1,
  pp. 1691--1700, 2024.

\bibitem{SAGIN1}
J.~Ye, S.~Dang, B.~Shihada, and M.-S. Alouini, ``Space-air-ground integrated
  networks: Outage performance analysis,'' \emph{IEEE Transactions on Wireless
  Communications}, vol.~19, no.~12, pp. 7897--7912, 2020.

\bibitem{SAGIN2}
J.~Tan, F.~Tang, M.~Zhao, and N.~Kato, ``Outage probability, performance, and
  fairness analysis of space-air-ground integrated network ({SAGIN}): Uav
  altitude and position angle,'' \emph{IEEE Transactions on Wireless
  Communications}, vol.~24, no.~2, pp. 940--954, 2025.

\bibitem{SAGIN3}
J.~Zhou, S.~Dang, B.~Shihada, and M.-S. Alouini, ``On the outage performance of
  space-air-ground integrated networks in the 3{D} poisson field,'' \emph{IEEE
  Transactions on Vehicular Technology}, vol.~73, no.~3, pp. 4401--4406, 2024.

\bibitem{SAGIN4}
Q.~Chen, W.~Meng, S.~Han, C.~Li, and T.~Q.~S. Quek, ``Coverage analysis of
  {SAGIN} with sectorized beam pattern under shadowed-rician fading channels,''
  \emph{IEEE Transactions on Communications}, vol.~71, no.~8, pp. 4988--5004,
  2023.

\bibitem{EC1}
H.~Niu, X.~Zhao, and J.~Li, ``3d location and resource allocation optimization
  for {UAV}-enabled emergency networks under statistical qos constraint,''
  \emph{IEEE Access}, vol.~9, pp. 41\,566--41\,576, 2021.

\bibitem{EC2}
J.~Wang, W.~Cheng, and H.~Zhang, ``Caching and d2d assisted wireless emergency
  communications networks with statistical qos provisioning,'' \emph{Journal of
  Communications and Information Networks}, vol.~5, no.~3, pp. 282--293, 2020.

\bibitem{myf}
Y.~Chen and W.~Cheng, ``Performance analysis of {RIS}-equipped-{UAV} based
  emergency wireless communications,'' in \emph{ICC 2022 - IEEE International
  Conference on Communications}, 2022, pp. 255--260.

\bibitem{Fchannel}
S.~K. Yoo, S.~L. Cotton, P.~C. Sofotasios, M.~Matthaiou, M.~Valkama, and G.~K.
  Karagiannidis, ``The {F}isher–{S}nedecor $\mathcal {F}$ distribution: A
  simple and accurate composite fading model,'' \emph{IEEE Communications
  Letters}, vol.~21, no.~7, pp. 1661--1664, 2017.

\bibitem{Fchannel3}
O.~S. Badarneh, D.~B. da~Costa, P.~C. Sofotasios, S.~Muhaidat, and S.~L.
  Cotton, ``On the sum of fisher\mbox{-}snedecor $\mathcal{F}$ variates and its
  application to maximal-ratio combining,'' \emph{IEEE Wireless Communications
  Letters}, vol.~7, no.~6, pp. 966--969, 2018.

\bibitem{tcomF}
Y.~Chen, W.~Cheng, and W.~Zhang, ``Reconfigurable intelligent surface equipped
  {UAV} in emergency wireless communications: A new fading–shadowing model
  and performance analysis,'' \emph{IEEE Transactions on Communications},
  vol.~72, no.~3, pp. 1821--1834, 2024.

\bibitem{shadowed1}
J.~Wang, W.~Cheng, and W.~Zhang, ``Performance boundary analyses for
  statistical multi-{QoS} framework over 6{G} sagins,'' \emph{IEEE Transactions
  on Wireless Communications}, pp. 1--1, 2025.

\bibitem{shadowed2}
E.~Kim, I.~P. Roberts, and J.~G. Andrews, ``Downlink analysis and evaluation of
  multi-beam leo satellite communication in shadowed rician channels,''
  \emph{IEEE Transactions on Vehicular Technology}, vol.~73, no.~2, pp.
  2061--2075, 2024.

\bibitem{shadowed3}
N.~I. Miridakis, D.~D. Vergados, and A.~Michalas, ``Dual-hop communication over
  a satellite relay and shadowed rician channels,'' \emph{IEEE Transactions on
  Vehicular Technology}, vol.~64, no.~9, pp. 4031--4040, 2015.

\bibitem{shadowed4}
D.-H. Jung, J.-G. Ryu, W.-J. Byun, and J.~Choi, ``Performance analysis of
  satellite communication system under the shadowed-rician fading: A stochastic
  geometry approach,'' \emph{IEEE Transactions on Communications}, vol.~70,
  no.~4, pp. 2707--2721, 2022.

\bibitem{2007Table}
I.~S. Gradshteyn and I.~M. Ryzhik, ``Table of integrals, series, and
  products,'' \emph{Mathematics of Computation}, vol.~20, no.~96, p.
  1157–1160, 2007.

\bibitem{Integrals}
A.~P. Prudnikov, Y.~A. Brychkov, and O.~I. Marichev, ``Integrals and series of
  elementary functions,'' \emph{Mathematics of Computation}, vol.~40, no. 161,
  1981.

\bibitem{LAP}
A.~Al-Hourani, S.~Kandeepan, and S.~Lardner, ``Optimal {LAP} altitude for
  maximum coverage,'' \emph{IEEE Wireless Communications Letters}, vol.~3,
  no.~6, pp. 569--572, 2014.

\bibitem{fixedgainAF}
M.~Hasna and M.-S. Alouini, ``A performance study of dual-hop transmissions
  with fixed gain relays,'' \emph{IEEE Transactions on Wireless
  Communications}, vol.~3, no.~6, pp. 1963--1968, 2004.

\bibitem{Variablegain_AF_DF}
Z.~Zhang, Q.~Sun, M.~López-Benítez, X.~Chen, and J.~Zhang, ``Performance
  analysis of dual-hop rf/fso relaying systems with imperfect csi,'' \emph{IEEE
  Transactions on Vehicular Technology}, vol.~71, no.~5, pp. 4965--4976, 2022.

\bibitem{walfram}
I.~Wolfram~Res., ``{T}he wolfram functions,'' \url{Available:
  http://functions.wolfram.com}, accessed: Mar. 25, 2024.[Online].

\bibitem{AF-YANG}
S.~Li, L.~Yang, D.~B. da~Costa, and S.~Yu, ``Performance analysis of uav-based
  mixed rf-uwoc transmission systems,'' \emph{IEEE Transactions on
  Communications}, vol.~69, no.~8, pp. 5559--5572, 2021.

\bibitem{AFasy}
H.~Lei, Y.~Zhang, K.~Park, I.~S. Ansari, and M.~S. Alouini, ``Performance
  analysis of dual-hop rf-uwoc systems,'' \emph{IEEE Photonics Journal},
  vol.~PP, no.~99, pp. 1--1, 2020.

\bibitem{DF}
B.~Ashrafzadeh, E.~Soleimani-Nasab, M.~Kamandar, and M.~Uysal, ``A framework on
  the performance analysis of dual-hop mixed {FSO-RF} cooperative systems,''
  \emph{IEEE Transactions on Communications}, vol.~67, no.~7, pp. 4939--4954,
  2019.

\bibitem{2007qos}
J.~Tang and X.~Zhang, ``Cross-layer resource allocation over wireless relay
  networks for quality of service provisioning,'' \emph{IEEE Journal on
  Selected Areas in Communications}, vol.~25, no.~4, pp. 645--656, 2007.

\bibitem{outagecapacity}
K.-L. Besser and E.~A. Jorswieck, ``Bounds on the outage probability in
  dependent rayleigh fading channels,'' in \emph{ICC 2020 - 2020 IEEE
  International Conference on Communications (ICC)}, 2020, pp. 1--6.

\bibitem{Bandwidth}
L.~Shi, L.~Zhao, G.~Zheng, Z.~Han, and Y.~Ye, ``Incentive design for
  cache-enabled d2d underlaid cellular networks using stackelberg game,''
  \emph{IEEE Transactions on Vehicular Technology}, vol.~68, no.~1, pp.
  765--779, 2019.

\bibitem{T}
W.~Cheng, X.~Zhang, and H.~Zhang, ``Statistical-{QoS} driven energy-efficiency
  optimization over green 5{G} mobile wireless networks,'' \emph{IEEE Journal
  on Selected Areas in Communications}, vol.~34, no.~12, pp. 3092--3107, 2016.

\bibitem{channelmeasurement}
N.~Bhargav, S.~L. Cotton, and D.~E. Simmons, ``Secrecy capacity analysis over
  $\kappa $--$\mu $ fading channels: Theory and applications,'' \emph{IEEE
  Transactions on Communications}, vol.~64, no.~7, pp. 3011--3024, 2016.

\end{thebibliography}
\end{document}